\documentclass[%
 reprint,
 amsmath,amssymb,
 aps,
longbibliography
]{revtex4-2}

\usepackage{graphicx}
\usepackage{dcolumn}
\usepackage{bm}
\PassOptionsToPackage{hyphens}{url}
\usepackage{hyperref} 
\hypersetup{colorlinks=true, linkcolor=blue, citecolor=ForestGreen, urlcolor=DarkOrchid}
\usepackage{dsfont}

\usepackage[normalem]{ulem}

\usepackage[dvipsnames]{xcolor}

\begin{document}

\title{Mean-Field Approximate Optimization Algorithm}

\author{{Aditi Misra-Spieldenner}$^{1}$, Tim Bode$^{2, 3}$,  Peter K. Schuhmacher$^{3}$, Tobias Stollenwerk$^{2, 3}$, Dmitry Bagrets$^{2, 4}$, and Frank K. Wilhelm$^{1, 2}$}
\affiliation{%
 $^{1}$Theoretical Physics, Saarland University, 66123 Saarbrücken, Germany\\
 $^{2}$Institute for Quantum Computing Analytics (PGI-12), Forschungszentrum Jülich, 52425 Jülich, Germany\\
 $^{3}$Institute for Software Technology, German Aerospace Center (DLR), Linder Höhe, 51147 Cologne, Germany\\
 $^{4}$Institute for Theoretical Physics, University of Cologne, 50937 Cologne, Germany
}%

\date{\today}

\begin{abstract}
The Quantum Approximate Optimization Algorithm (QAOA) is suggested as a promising application on early quantum computers. Here, a quantum-inspired classical algorithm, the {mean-field} Approximate Optimization Algorithm (mean-field AOA), is developed by replacing the quantum evolution of the QAOA with classical spin dynamics through the mean-field approximation. Due to the alternating structure of the QAOA, this classical dynamics can be found exactly for any number of QAOA layers. We benchmark its performance against the QAOA on the Sherrington-Kirkpatrick (SK) model and the partition problem, and find that the mean-field AOA outperforms the QAOA in both cases for most instances. Our algorithm can thus serve as a tool to delineate optimization problems that can be solved classically from those that cannot, i.e.\ we believe that it will help to identify instances where a true quantum advantage can be expected from the QAOA. To quantify quantum fluctuations around the mean-field trajectories, we solve an effective scattering problem in time, which is characterized by a spectrum of time-dependent Lyapunov exponents. These provide an indicator for the hardness of a given optimization problem relative to the mean-field AOA.
\end{abstract}

\maketitle


\section{\label{sec:intro}Introduction}

A large class of NP-hard optimization problems admits a formulation as an Ising model, such that the optimum corresponds to the ground state while the hardness is related to the spin-glass phase of the Ising Hamiltonian~\cite{Lucas:2014}.
{A potentially powerful strategy to find this desired ground state is {adiabatic quantum computation} (AQC)~\cite{Apolloni1989,Apolloni1990,Kadowaki1998,Farhi2001,Bapst:2013,Lidar_2018}.
Here, a spin system is initialized in the unique and easily accessible ground state of a {driver} Hamiltonian $\hat{H}_{D}$ and afterwards transferred  {adiabatically} to the desired Ising problem Hamiltonian $\hat H_{P}$.
The adiabatic theorem then guarantees that the spin system remains in its instantaneous ground state throughout the entire time evolution, and in particular at the final time when reaching the problem Hamiltonian.
To keep the evolution  {truly} adiabatic, however, the sweep velocity has to be carefully chosen as a function of the minimal gap between the instantaneous ground and first excited states.
Unfortunately, this minimal gap becomes exponentially small for typical hard instances, forcing the evolution time to become exponentially large \cite{VanDam2001,Reichardt2004,Amin2009,Altshuler2010}.} {Inspired by} AQC, the {Quantum Approximate Optimization Algorithm} (QAOA) for solving this type of combinatorial problems was suggested as a diabatic alternative~\cite{Farhi:2014}: {The time evolution of a linear annealing schedule is discretized using the standard Suzuki-Trotter decomposition, such that it becomes an alternating sequence of parameterized unitary gates of some given length $p$, applied to the initial state. Note that for noisy intermediate-scale quantum (NISQ) devices, the number of layers $p$ is naturally limited~\cite{Preskill2018}. Hence, instead of using the linear annealing schedule for some large $p$ as in AQC, the parameters of the unitaries are  {optimized} in a closed loop such that the energy expectation value of the problem Hamiltonian becomes minimal at the end of the circuit. Due to the heuristic nature of the QAOA, its actual computational power remains unclear up to date. In particular, the question arises for which kind of problems a quantum advantage relative to classical optimization algorithms can be expected \cite{Farhi:2016,Crooks2018,Hadfield2019,Zhou2020,Dupont2022}.

In this work, we present a {classical} algorithm inspired by the QAOA, called  {mean-field Approximate Optimization Algorithm} (mean-field AOA). The new algorithm replaces the quantum evolution of the QAOA by {classical} spin dynamics. Within the standard Trotterization scheme of the QAOA~\cite{Brady_2021}, this dynamics can be found  {exactly} for any number of layers $p$.

The algorithm can serve as an additional tool to delineate optimization problems that can be solved classically from those that cannot, i.e.\ we believe that it will help to identify instances where a true quantum advantage can be expected from the QAOA. Also, for instances where the mean-field AOA delivers a good solution, one can expect that such an advantage is not forthcoming. 

By introducing a path-integral representation based on spin-coherent states~\cite{Stone:2000}, we prove that for $p\gg 1$ the mean-field AOA emerges as an approximation to QAOA that is applicable for a large number of spins $N$ and a large average degree of the underlying problem graphs.
Performance tests of the mean-field AOA on two benchmark problems  -- the  {Sherrington-Kirkpatrick} (SK) model and the  number {partition problem} (NP) --- suggest that it delivers an approximate optimum with accuracy of order $1/N^{\delta}$ in polynomial time and thus outperforms the QAOA on average
(here $\delta>0$ is a problem specific exponent which is equal to $1/4$ and $0.95$ for the SK model and the NP problem, respectively). 

To quantify quantum fluctuations around the mean-field spin trajectories, we solve an effective scattering problem in time. It describes the propagation of collective `paramagnon' modes above the instantaneous ground state of the adiabatic Hamiltonian, and is characterized by a spectrum of positive Lyapunov exponents. The largest Lyapunov exponent shows a number of maxima, 
which are pinned to level crossings or minimal gaps in the lowest part of the Hamiltonian spectrum.
The case where all Lyapunov exponents are relatively small indicates an `easy' instance, where the optimum can be found classically, i.e.\ without invoking quantum algorithms. For hard instances, there occur some large maxima in the spectrum, and the mean-field AOA typically fails to deliver the exact solution.

It is known from studies of the SK and related Hopfield models that the first occurrence of the mini-gap in the course of the adiabatic protocol is related to the ergodic-to-MBL (many-body localization) transition in the spectrum of the corresponding quantum 
adiabatic Hamiltonian~\cite{Knysh2016, Mukherjee:2018, Alet:2018, Wang2022}.  Therefore, as a by-product, our fluctuation analysis enables one to approximately locate the instance-specific critical 
points of these transitions, essentially without the need of an expensive exact diagonalization which becomes out of reach for large system sizes.

This article is organized as follows.
In section \ref{sec:mf_aoa}, we begin by introducing the mean-field AOA, followed by an illustration of its performance for the SK model and the partition problem. Next, in section \ref{sec:path_int}, we discuss the spin coherent-state path integral, the saddle-point of which corresponds to our classical algorithm. Here, we also study the Gaussian quantum fluctuations around the classical path, characterized by the spectrum of Lyapunov exponents. Finally, in section \ref{sec:discussion_outlook}, we discuss possible future directions. Note that an implementation of the numerical code used in this paper, alongside our research data, is available online~\cite{mean_field_code}. A corresponding software package in Julia~\cite{bezanson2017julia} has also been implemented~\cite{QAOA_jl}.

\section{\label{sec:mf_aoa}Mean-Field Approximate Optimization Algorithm}

Throughout this article, we adopt the following form of the problem and driving Hamiltonians 
\begin{subequations}\label{eq:H}
\begin{align}
\hat{H}_{P} &= - \sum_{i=1}^N \bigg[ h_i  + \sum_{j>i} J_{ij}  \hat\sigma^z_j \bigg] \hat\sigma^z_i, \label{eq:H_P}\\
\hat{H}_{D} &= - \sum_{i=1}^N \Delta_i \hat\sigma^x_i, \quad  \Delta_i>0 \label{eq:H_D}.
\end{align}
\end{subequations}
The positivity of all constants $\Delta_i$ guarantees that the $N$-qubit state
\begin{align}\begin{split}
\label{eq:psi_in}
    |\psi_0\rangle = |+\rangle_1^X \otimes |+\rangle_2^X \otimes \cdots \otimes |+\rangle_N^X
\end{split}\end{align}
is the ground state of $\hat{H}_{D}$.
For the standard QAOA, one usually sets $\Delta_i = 1$, such that this driving frequency becomes a natural choice as frequency unit and inverse timescale, {which we adopt unless otherwise stated}.
Note also that $\hat{H}_{P}$ with {positive} couplings $J_{ij}>0$  and vanishing local fields $h_i \to 0$ describes the {ferromagnetic} state of the Ising model.
Our choice of numerical factors and signs is in correspondence to Ref.~\cite{Lucas:2014}. 

Inspired by the alternating application of $\hat{H}_{D}$ and $\hat{H}_{P}$ in the standard QAOA, we derive the mean-field equations of motion for two separate cases in the following: (i) only the driving Hamiltonian $\hat{H}_{D}$ is active, (ii) only the problem Hamiltonian $\hat{H}_{P}$ is active. 

In the mean-field approximation, the total Hamiltonian then becomes 
\begin{align}
\label{eq:H_classical}
    \begin{split}
        H(t) ={} &- \gamma(t)\sum_{i=1}^N \bigg[ h_i + \sum_{j>i} J_{ij}  n_j^z(t) \bigg] n_i^z(t) \\
        &- \beta(t) \sum_{i=1}^N \Delta_i n_i^x(t),
    \end{split}
\end{align}
where $\beta(t)$ and $\gamma(t)$ are piecewise-constant functions of time, and we assume $J_{ii} = 0$ without loss of generality. The classical spin vectors are defined as
\begin{align}
    \boldsymbol{n}_i(t) &= \left(n_i^x(t), n_i^y(t), n_i^z(t)\right)^T \\
    &= \left(\operatorname{Tr}\left[\hat{\rho}\,\hat\sigma^x_i(t)\right], \operatorname{Tr}\left[\hat{\rho}\,\hat\sigma^y_{i}(t)\right], \operatorname{Tr}\left[\hat{\rho}\,\hat\sigma^z_i(t)\right]\right)^T, \nonumber
\end{align}
where the system density matrix factorizes in the mean-field approximation,
\begin{align}
    \hat{\rho} &= \bigotimes_{i=1}^{N}\hat{\rho}^{(i)}.
\end{align}
Further details on this approximation are provided in Appendix~\ref{app:QAOA_vs_MF}. We also introduce the effective magnetization
\begin{align}\begin{split}\label{eq:magnetization}
m_i(t) = h_i + \sum_{j=1}^N J_{ij} n_j^z(t).
\end{split}\end{align}
The dynamics of the system then amounts to a precession of each spin in an effective magnetic field, i.e.\
\begin{align}\begin{split}\label{eq:eom_mf}
    \partial_t \boldsymbol{n}_i(t) = \boldsymbol{n}_i(t) \times \boldsymbol{B}_i(t),
\end{split}\end{align}
where $\boldsymbol{B}_i(t) = 2\beta(t)\Delta_i \boldsymbol{\hat{e}}_x + 2\gamma(t)m_i(t) \boldsymbol{\hat{e}}_z$. This leads to 
\begin{align}\begin{split}
\dot n_i^y(t) = 2 \Delta_i  n_i^z(t), \quad \dot n_i^z(t) = - 2 \Delta_i  n_i^y(t),
\end{split}\end{align}
and $n_i^x(t) = n_i^x(0)$ 
during the time intervals with $\beta(t)\equiv 1$, $\gamma(t)\equiv 0$ corresponding to case (i), while we have
\begin{align}\begin{split}
\dot n_i^x(t) =   2 m_i(t) n_i^y, \quad \dot n_i^y(t) =  -2 m_i(t) n_i^x(t)
\end{split}\end{align}
and $n_i^z(t) = n_i^z(0)$ in the complementary case (ii) where $\gamma(t)\equiv 1$, $\beta(t)\equiv 0$.
The norm of all spin vectors is conserved under this unitary evolution, $\left|\boldsymbol{n}_i(t)\right|^2 = 1$.

Solving the above differential equations for the typical piecewise constant Hamiltonian governing the QAOA, one obtains after $p$ iterations:
\begin{align}\begin{split}\label{eq:n(t)}
\boldsymbol{n}_i(p) = \prod_{k=1}^p \hat V_i^D(k) \hat V_i^P(k) \boldsymbol{n}_i(0).
\end{split}\end{align}
In Eq.~\eqref{eq:n(t)}, the two unitary matrices $3\times 3$ are defined as
\begin{align}\begin{split}\label{eq:V_D}
\hat V_i^D(k) = 
\begin{pmatrix}
1 & 0 & 0 \\
0 & \phantom{-}\cos(2 \Delta_i\beta_k) & \sin(2 \Delta_i \beta_k) \\
0 & -\sin (2 \Delta_i \beta_k) & \cos(2 \Delta_i \beta_k) 
\end{pmatrix}
\end{split}\end{align}
and
\begin{align}\begin{split}\label{eq:V_P}
\hat V_i^P(k) = 
\begin{pmatrix}
\phantom{-}\cos(2m_i (t_{k-1}) \gamma_k) & \sin(2m_i (t_{k-1}) \gamma_k) & 0 \\
-\sin (2m_i (t_{k-1}) \gamma_k) & \cos(2m_i (t_{k-1}) \gamma_k) & 0 \\
0 & 0 & 1
\end{pmatrix},
\end{split}\end{align}
where $t_k = k \tau$.
For the parameters $\beta_k$ and $\gamma_k$, which are conjugate to $\hat{H}_{D}$ and $\hat{H}_{P}$, respectively, it is sufficient to take linear functions inspired by the adiabatic quantum algorithm~\cite{Willsch2020}.
For $k=1, ..., p$ we then have
\begin{equation}
\label{eq:adiabatic}
\gamma_k = \tau k/p, \qquad \beta_k = \tau \left(1 - (k-1)/p\right), 
\end{equation} 
where $\tau$ is the time step which should be adjusted so that the the spin dynamics remains regular. Unless otherwise stated, for our algorithm we often resort to using $p \sim 10^4$ and $\tau \sim 1/2$. 

\subsection{Algorithm}
\label{subsec:mf_AOA}

The obtained analytical expressions for the time evolution of the classical spin vector for each of the $N$ qubits under the mean-field approximation will now be employed to create a quantum-inspired {classical} algorithm which we call {mean-field} AOA. To apply the algorithm, the following steps have to be completed \footnote{Note that if the Hamiltonian $\hat{H}_{P}$ possesses $\mathcal{Z}_2$ symmetry, i.e.\ $h_i=0$ for all $i=1, ..., N$, then it is crucial to explicitly {break} this symmetry by fixing one of the spins (we typically fix the `last' spin to $+1$, thus introducing local magnetic fields as in Eq.~\eqref{eq:h_spin_fix}). Otherwise, the algorithm will simply remain in the initial state.}:  
\begin{enumerate}
    \item Initialize the $N$ classical spin vectors in the state
    \begin{align}
        \boldsymbol{n}_i(0) = (1, 0, 0)^T \quad \forall i.
    \end{align}
    This is analogous to the uniform superposition of all computational input states used in the QAOA (other initial states are also possible).
    
    \item Apply a mean-field evolution sequence of length $p\in\mathbb{N}$ to $\boldsymbol{n}_i(0)$ such that
    \begin{align}\begin{split}
    \label{eq:n_i_p}
        \boldsymbol{n}_i(p) = \prod_{k=1}^p \hat V_i^D(k) \hat V_i^P(k) \boldsymbol{n}_i(0),
    \end{split}\end{align}
    where $\hat V_i^D(k)$ and $\hat V_i^P(k)$ are given in Eqs.~\eqref{eq:V_D} and \eqref{eq:V_P}, respectively.
Observe that $\hat V_i^P(k)$ depends on $\boldsymbol{n}_i(k-1)$.
The schedule for the parameters $\beta_k$, $\gamma_k$ is given in Eq.~\eqref{eq:adiabatic}.

    \item Compute the cost function
    \begin{align}
       {H}_{P}[{p, \tau}] &= - \sum_{i=1}^N \bigg[ h_i + \sum_{j>i} J_{ij}  n_j^z(p) \bigg] n_i^z(p).
    \end{align}

    \item Adjust the number of steps $p\in\mathbb{N}$ and the step size $\tau$  to minimize the cost function.
    
    \item Repeat steps 2.
    to 4. until a convergence threshold is reached.

    \item {Round the $z$-components of the spin vectors $\boldsymbol{n}_i(p)$ to obtain the resulting bitstring
    \begin{align}
        \boldsymbol{\sigma}_* = \left(\mathrm{sign}(n_1^z), ..., \mathrm{sign}(n_N^z) \right).
    \end{align}}
\end{enumerate}

Note that our formulation of the mean-field AOA deliberately does not include an optimization of the cost function over the parameters $\beta_k$ and $\gamma_k$. As we will show in section \ref{subsec:performance}, the advantage of our classical algorithm is that it is not limited to a small number of steps $p$, which allows us to perform adiabatically slow evolution with the annealing schedule defined in Eq.~\eqref{eq:adiabatic}. 

Regarding the fourth step of our algorithm, we point out that the returned solution bitstrings $\boldsymbol{\sigma}_*$ are very robust with respect to changes in $\tau$ and $p$, i.e.\ the final magnetic orientation of each spin is largely determined by the structure of the classical phase space. In practice, it suffices to start with a relatively large step size $\tau = 1/2$ and, for example, $p = 10^3$. Subsequently, $\tau$ can be decreased, if necessary, while $p$ should be increased until a set of smooth trajectories is reached and the final solution remains unaltered.

We close this subsection by two important comments. First, the outlined algorithm is polynomial in time and scales as $O(p N^2)$. Namely, for a given step $k=1, \dots, p$, see Eq.~(\ref{eq:n_i_p}}), and for every spin $i=1, ..., N$, one needs to perform the sum for the magnetization $m_i(t_k)$ in Eq.~(\ref{eq:magnetization}), which leads to an additional factor of $N$. Second, the dynamics of spins described by the system of non-linear differential equations~(\ref{eq:eom_mf}) is in general \textit{chaotic} for random optimization problems. Our algorithm explores a small fraction of the phase space with energy $H(t)$ that lies close to the edge of the spectrum of
the adiabatic Hamiltonian $H(s) = (1-s) H_D + s H_P$. The classical dynamics in this region happens to be regular provided a small enough step size $\tau < \tau_c$ is chosen, 
where $\tau_c$ marks the transition point to the chaotic regime. We have found the critical $\tau_c$ to be of order unity for both the SK model and the partition problem, as analyzed below in more detail. Contrary to this, if one starts from an \textit{excited} state, e.g.\ by flipping at least one of the spins in the initial state to $(-1,0,0)^T$, then the subsequent dynamics shows chaotic behavior for \textit{any} step size~$\tau$.

\subsection{Performance}\label{subsec:performance}

The performance of the algorithm outlined above will now be tested by comparing its results to those of the standard QAOA.
As an introductory problem, in section \ref{subsubsec:SK_model} we investigate the SK model~\cite{SK_1975, panchenko2013sherrington} well-known from disordered systems and spin glasses.
Subsequently, in section~\ref{subsubsec:number_part}, we analyze the {partition problem}, i.e.\ the problem of partitioning a set of positive integers into two subsets such that their respective sums are as close to equal as possible.
In both cases, we provide numerical evidence that the large-$N$ scaling of the mean-field AOA consistently outperforms that of the QAOA for finite $p$.

\subsubsection{Sherrington-Kirkpatrick Model}\label{subsubsec:SK_model}

\begin{figure}[t!]
	\begin{center}
		\includegraphics[width=\columnwidth]{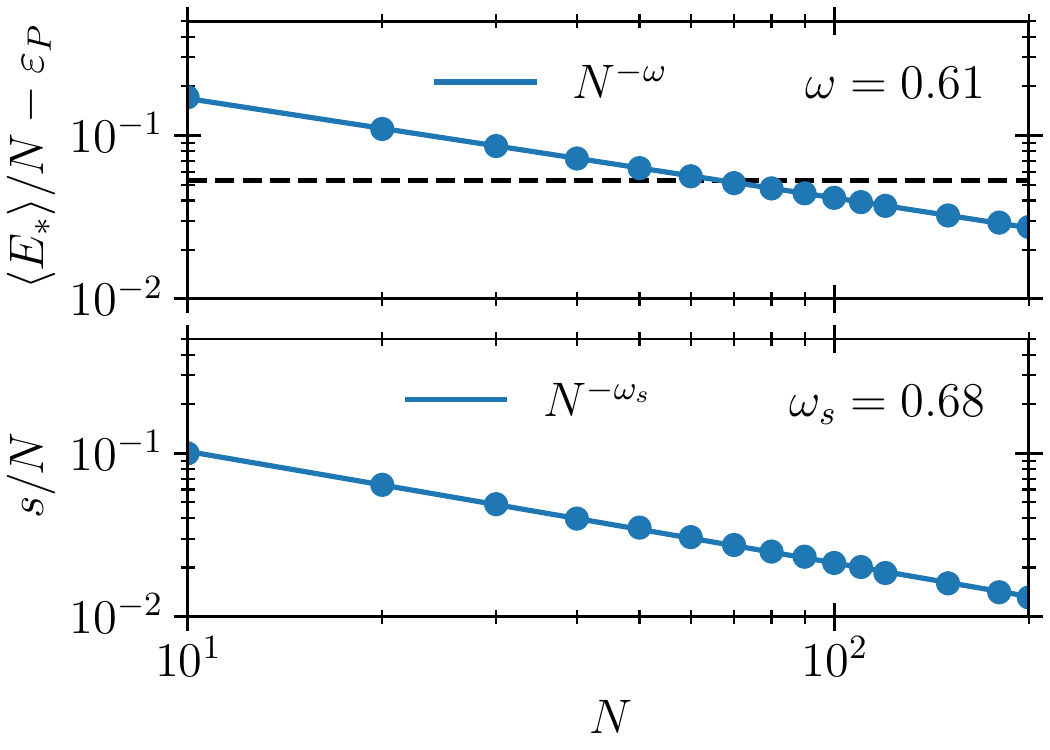}
		\caption
		{\label{fig:Algebraic_decrement_SK} Scaling of the energy $E_*$ resulting from our algorithm with the system size, as estimated for $10^4$ random instances of the SK model. $\varepsilon_P=-0.763$ is the Parisi value. The dashed line shows the lowest energy achievable with zero-temperature annealing, quoted as $-0.71$ in Ref.\ \cite{Farhi:2022}. The value $\omega = 0.61$ is to be compared with the value $\omega = 2/3$ from Ref.~\cite{Palassini_2008}. In the lower panel, we show the estimated standard deviation with corresponding scaling. The parameters of the schedule in Eq.~\eqref{eq:adiabatic} are chosen throughout as $\tau = 1/2$ and $p = 10^3$. The data point to the very right is $N=200$.}
	\end{center}
\end{figure}

The SK model was first introduced in the context of spin glasses but can also be understood more broadly as an optimization problem where $N$ coupled spins are to be distributed into two subgroups according to the sign of their coupling~\cite{panchenko2013sherrington}.
In our convention, the goal is then to {minimize} the (classical) cost function
\begin{align}\begin{split}\label{eq:H_SK}
    \hat H_P = - \frac{1}{\sqrt{N}}\sum_{i<j\leq N} J_{ij} \hat\sigma^z_i \hat\sigma^z_j,
\end{split}\end{align} 
where the couplings $J_{ij}$ are i.i.d.\ standard Gaussian variables, i.e.\ with zero mean $\left\langle J_{ij} \right\rangle = 0$ and variance $ \left\langle J_{ij}^2 \right\rangle = J^2$.
Note that in the limit $N\to\infty$, mean-field theory becomes exact for this model~\cite{SK_1975}.

To remove the degeneracy caused by the $\mathcal{Z}_2$ symmetry of Eq.~\eqref{eq:H_SK}, we fix the `final' spin to be in {the state $\hat{\sigma}^z_N \left|0\right\rangle_N = \left|0\right\rangle_N$}.
This leads to an equivalent cost function in the form of Eq.~\eqref{eq:H_P}
with random local magnetic fields
\begin{align}\label{eq:h_spin_fix}
    h_i = J_{iN}, \quad i=1, \dots N-1.
\end{align}
In the thermodynamic limit, the SK model is also known~\cite{Parisi_1979} to have the ground-state energy $E_0$ that on average converges to
\begin{align}\begin{split}
    \varepsilon_P = \lim_{N\to\infty}\left\langle E_0/N \right\rangle_J = - 0.763...
\end{split}\end{align} 
This theoretical value can be used to test the performance of both the mean-field algorithm and the QAOA, the latter having been performed recently in Ref.~\cite{Farhi:2022}. There it was found that the {QAOA} can surpass certain classical algorithms such as spectral relaxation~\cite{aizenman1987some} and semidefinite programming~\cite{10.1145/2897518.2897548} in the limit $N\to\infty$ at finite $p=12$. The energy benchmark from these classical algorithms is $\lim_{N\to\infty}\left\langle E_0/N \right\rangle_J \approx -2/\pi$. A classical algorithm capable of returning a result within an arbitrary distance from the optimum is given in Ref.~\cite{Montanari:2019}. Note also that the optimal variational parameters of the QAOA are found to be independent of the particular instance of the SK model~\cite{Farhi:2022}, i.e.\ one global schedule of parameters works best for all random instances.

In Fig.~\ref{fig:Algebraic_decrement_SK} we show our results for the approximate solution 
\begin{equation}
    E_* = \langle \hat H_P \rangle|_{\boldsymbol{\sigma}_*}
\end{equation}
as a function of the number of spins $N$. The ensemble average is taken over $10^4$ random instances of the SK model. The scaling of $\langle E_*/N\rangle$ with $N$ demonstrates that our algorithm outperforms both zero-temperature annealing (dashed line at $\sim \varepsilon_P-0.71$) and the QAOA at $p=12$ (cf. Ref.~\cite{Farhi:2022}, which shows that the latter beats the quoted value of $-2/\pi$). The scaling exponent is also in decent agreement with the results of previous very detailed numerical investigations of the SK model~\cite{Palassini_2008}.

\begin{figure}[t!]
	\begin{center}
		\includegraphics[width=.45\textwidth]{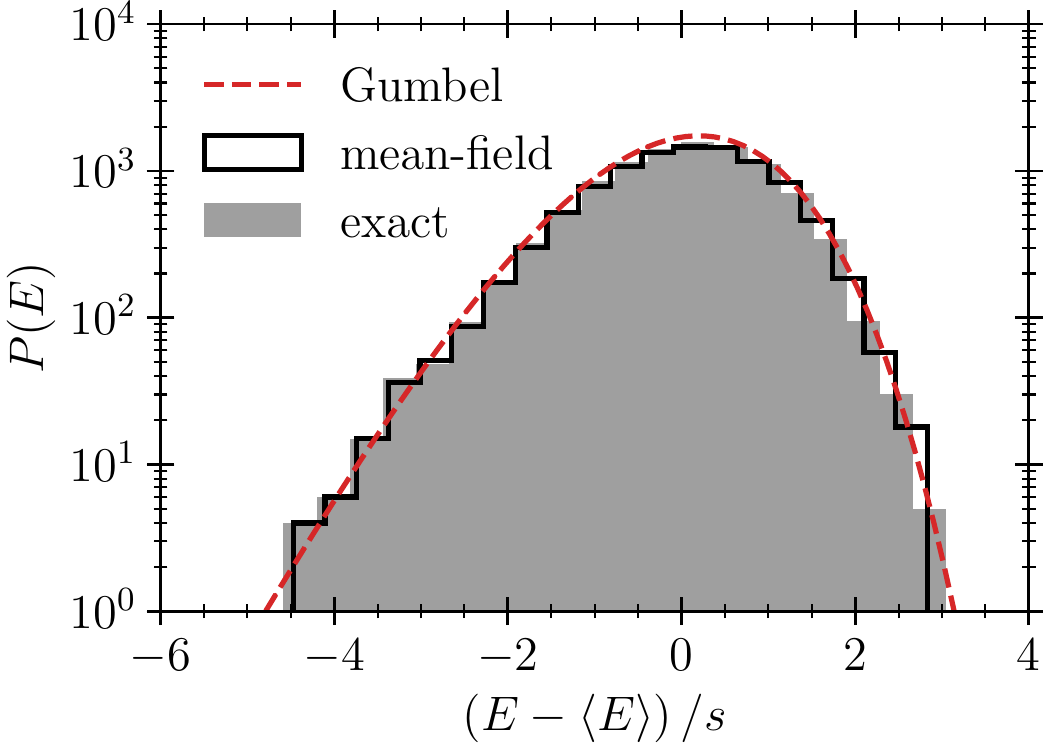}
		\caption
		{\label{fig:Gumbel_true_minima_SK} Comparison of the distribution of the exact ground state and the result of our algorithm for $10^4$ random instances of the SK model with $N=20$. The estimated standard deviation is again denoted by $s$. The (red) dashed line shows the Gumbel distribution Eq.~\eqref{eq:gumbel} for the $m$th smallest value as employed in Ref.~\cite{Palassini_2008}. Note that up to normalization, we use the same parameters as in \cite{Palassini_2008}, i.e.\ $m=6$, location $u=0.2$ and scale $v=2.35$.}
	\end{center}
\end{figure}

Instead of computing averages over the ensemble of $10^4$ instances, Fig.~\ref{fig:Gumbel_true_minima_SK} shows the distribution of the (exact) solutions for the particular case of $N=20$. The mean-field algorithm performs very well in comparison to the exact results. The solutions returned by the mean-field AOA  follow the Gumbel distribution for the $m$th smallest element with $m=6$~\cite{Palassini_2008}, which is defined as
\begin{align}\label{eq:gumbel}
    g_m(x)=w \exp \left[m \frac{x-u}{v}-m \exp\left(\frac{x-u}{v}\right) \right],
\end{align}
where $u,v$ are parameters defining the mean value and variance, while $w$ is a normalization constant. We found that the outcomes of our algorithm follow this distribution irrespective of the value of $N$. 

\begin{figure}[t!]
	\begin{center}
		\includegraphics[width=.45\textwidth]{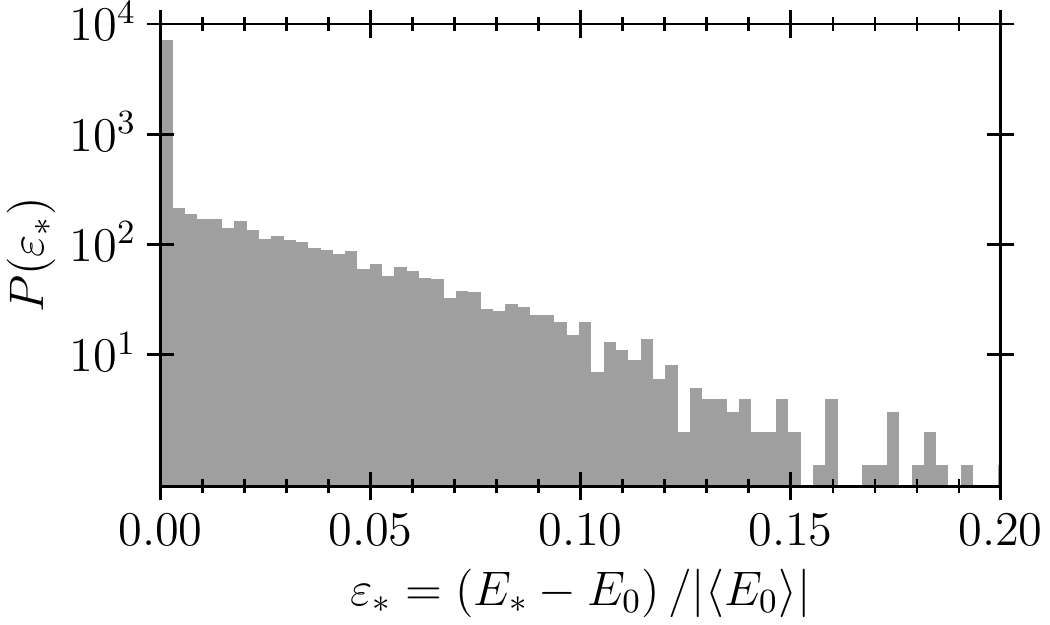}
		\caption
		{\label{fig:Tail_distribution_SK} Probability distribution of the minima returned by the mean-field AOA for $10^4$ random instances of the SK model at $N=20$. Note that the (blue) squares of Fig.~\ref{fig:Tail_distribution_Performance_SK} show the result of integrating the tail of this distribution from a given threshold $\varepsilon$ to infinity.
		}
	\end{center}
\end{figure}

\begin{figure}[t!]
	\begin{center}
		\includegraphics[width=.45\textwidth]{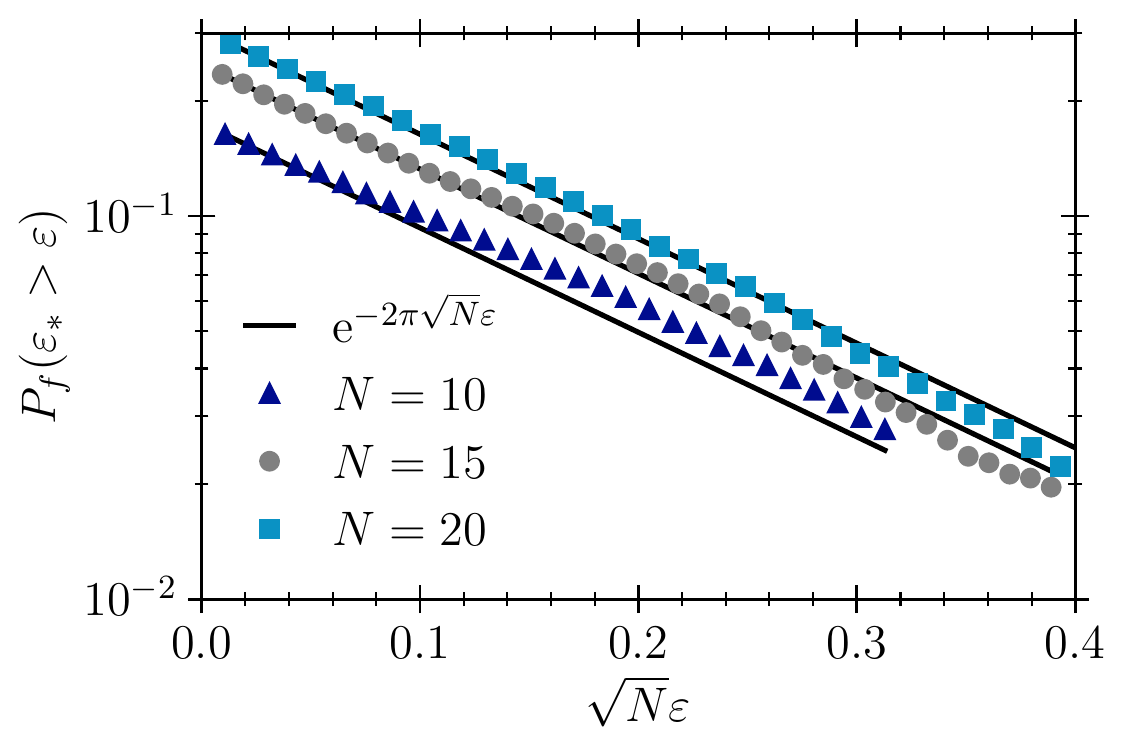}
		\caption
		{\label{fig:Tail_distribution_Performance_SK} Illustration of the scaling of the tail probability $P_f$ in Eq.~\eqref{eq:def_tail_prob}. The relation $\ln P_f \sim {-2\pi\sqrt{N} \varepsilon}$ is found to hold approximately for all $N\in\mathds{N}$ in the verified range $5, ..., 20$. For the SK model, the mean-field AOA thus returns an approximate local minimum that is algebraically close to the global one with almost unit probability.
		}
	\end{center}
\end{figure}

In Fig.~\ref{fig:Tail_distribution_SK}, we plot the success probability distribution of the classical algorithm, $P(\varepsilon_*)$, to deliver an approximate optimum at the relative distance $\varepsilon_* := (E_* - E_0) / \left|\langle E_0 \rangle\right|$ from the instance-specific true minimum $E_0$ (note the logarithmic scale for $P$). We then define the corresponding tail distribution
\begin{align}\label{eq:def_tail_prob}
    P_f(\varepsilon_* > \varepsilon) = \int_\varepsilon^\infty d\varepsilon_* P(\varepsilon_*), 
\end{align}
where $\varepsilon > 0$ is an arbitrary threshold. As exemplified in Fig.~\ref{fig:Tail_distribution_Performance_SK} for three different values of $N$, we find the failure probability to approximately follow the exponential law  
\begin{align}
\label{eq:P_f_SK}
    P_f(\varepsilon) \sim \exp{\left(-2\pi\sqrt{N} \varepsilon \right)},
\end{align}
which describes the statistics of rare events where our algorithm converges to high excited levels far from the ground state $E_0$. 
For large values of $N$, one may then pick the threshold to be $\varepsilon = N^{-1/4} \ll 1$ such that $P_f$ becomes exponentially small.
We then arrive at the following main conclusion of this section:
\begin{enumerate}
\item[]
{\it With a probability of at least $1 - O(\exp{(- 2\pi N^{1/4}}))$ over possible realizations of the SK Hamiltonian, the mean-field AOA delivers an approximate optimum $E_*$ with a relative accuracy bounded by $N^{-1/4}$ from above.
}
\end{enumerate}
\noindent In other words, in the limit $N\to\infty$, we find that the algorithm converges to the approximate solution $E_*$, which has an accuracy of at least $\varepsilon_*=N^{-1/4}$ almost with certainty. 
Since this analysis requires knowledge of the computationally expensive exact solutions $E_0$, we have restricted it to the range $N=5, ..., 20$.

\subsubsection{Number Partitioning}\label{subsubsec:number_part}

The partition problem for a set of natural numbers $\mathcal{S} = \{a_1, ..., a_N\} \subset \mathds{N}$ consists of finding two subsets $ \mathcal{S}_{1} \cup \mathcal{S}_2 =  \mathcal{S}$ such that the difference of the sums over the two subsets $\mathcal{S}_{1, 2}$ is as small as possible.
It belongs to the class of NP-complete problems~\cite{mezard2009information}.
The cost function can be written as
\begin{align}\begin{split}
    C(\mathcal{S}) = \sum_{i=1}^{N} a_i \hat\sigma^z_i.
\end{split}\end{align}
A so-called {perfect} partition occurs when $ C(\mathcal{S}) = \left(\sum_{a_i\in\mathcal{S}}a_i\right) \mod 2$. For bounded problems, the elements $a_i\in\mathcal{S}$ satisfy $1 \leq a_i \leq 2^M$ for some $M\in\mathds{N}$.

\begin{figure}[t!]
	\begin{center}
		\includegraphics[width=.45\textwidth]{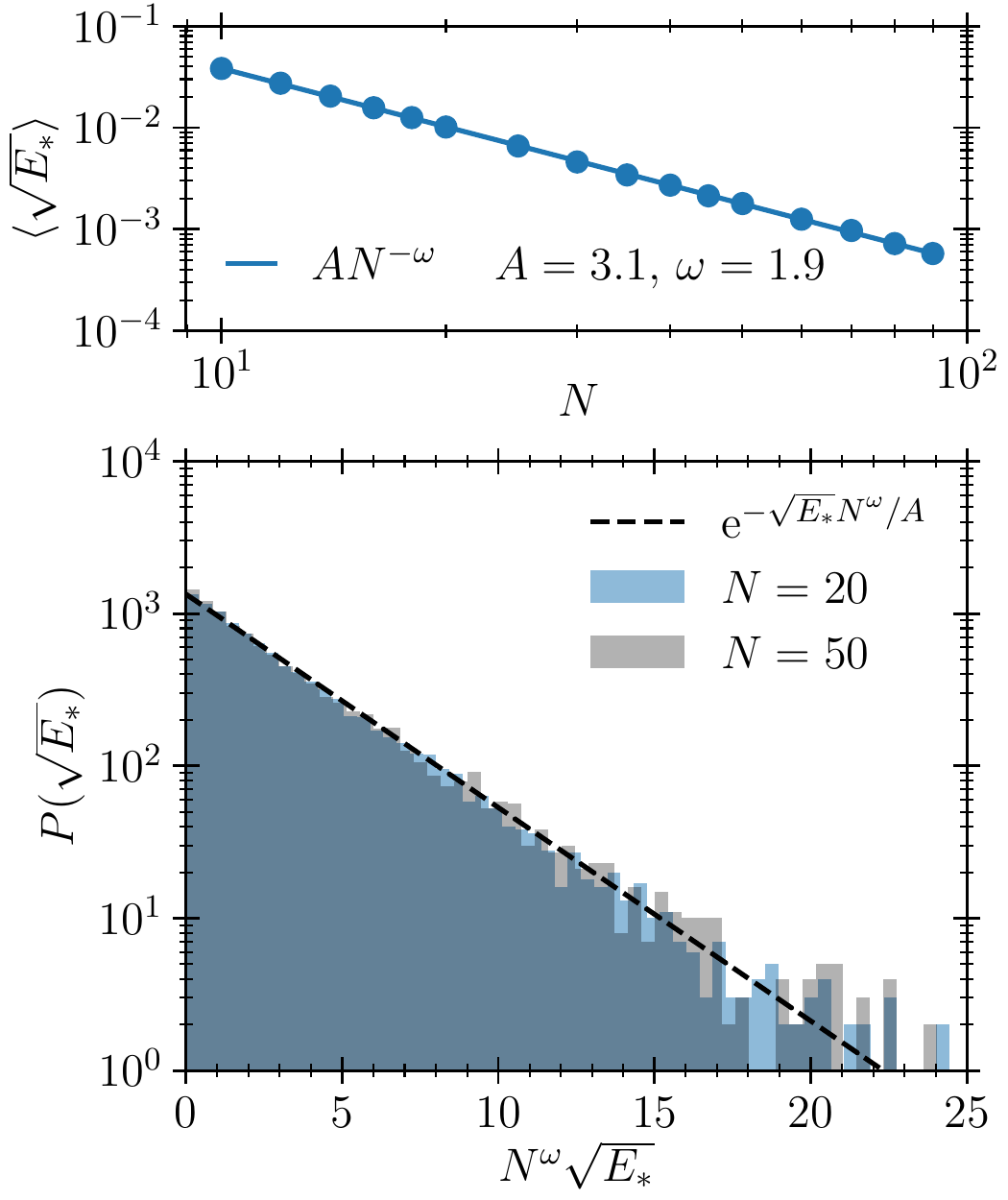}
		\caption
		{\label{fig:NP_exp_distro} Averaging over $10^4$ random realizations, we find the approximate cost $\sqrt{E_*}$ of the partition problem (as obtained from the mean-field AOA) to follow an exponential distribution. In the upper panel, we plot the first moment of this distribution as a function of the system size $N$ and extract the parameters $\omega$ and $A$. The lower panel shows the corresponding distribution. The parameters of the schedule in Eq.~\eqref{eq:adiabatic} are now $\tau = 1/4$ and $p = 10^4$.}
	\end{center}
\end{figure} 

\begin{figure}[t!]
	\begin{center}
		\includegraphics[width=.45\textwidth]{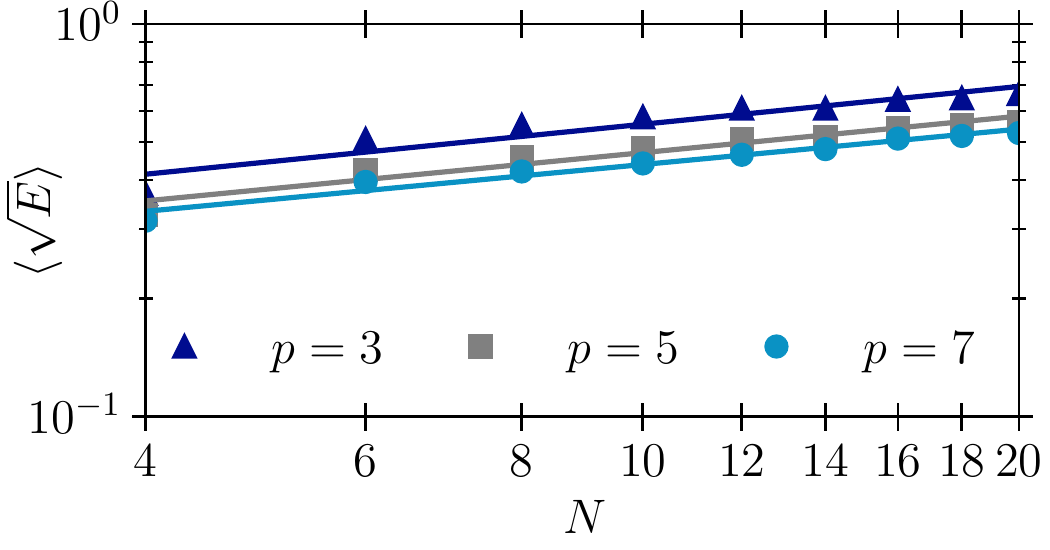}
		\caption
		{\label{fig:NP_QAOA} Average of the cost function over $10^2$ random realizations calculated from the energy expectation value of the simulated QAOA for each realization~\cite{bezanson2017julia, YaoFramework2019, Zygote}. The solid lines show a fit to $N^{\omega'}$, where in this case 
        ${\omega'} \approx 0.31 \pm 0.01$.}
	\end{center}
\end{figure} 

We focus on the case $M > N$, for which the large-$N$ {minimal} cost is expected to scale as $\sqrt{N}2^{M-N}$~\cite{mezard2009information}. In the thermodynamic limit (where both $M$ and $N$ go to infinity at some fixed ratio), the partition problem is known to have a phase transition at $M=N$. Numerically, almost all instances then cross over from having a perfect partition to having none. Note that neither our algorithm nor finite-$p$ QAOA simulations are able to resolve this satisfiability threshold.

The cost function $C(\mathcal{S})$ can be transformed to an Ising Hamiltonian by taking its square, i.e.\
\begin{align}\begin{split}
\label{eq:H_NP}
    \hat{H}_{P} &= C^2(\mathcal{S}) = \sum_{i=1}^N a_i^2 - \sum_{i<j\leq N} J_{ij} \hat\sigma^z_i \hat\sigma^z_j, 
\end{split}\end{align}
where we have introduced the couplings $J_{ij}=-2a_i a_j$. As before, we fix the `final' spin to be in {the state $\hat{\sigma}^z_N \left|0\right\rangle_N = \left|0\right\rangle_N$}.
To simplify the comparison of the mean-field and the quantum AOA, instead of natural numbers $a_i\in\mathds{N}$ we now take 
$a_i$ to be uniformly distributed in the unit interval $[0,1]$.
This is equivalent to a bounded problem with large $M$ where the $a_i$ are rescaled by $2^M$. Note that for double precision on a standard computer, one should thus effectively have $M=52$. It then follows that the induced distribution of couplings $J_{ij}$ is logarithmic,
\begin{align}
    P(J_{ij}) = \frac 1 2 \ln\left(-\frac{2}{J_{ij}}\right), \quad -2<J_{ij}<0.
\end{align}
In contrast to the SK model, there are non-vanishing correlations between $J_{ik}$ and $J_{il}$ for $l \neq k$.
A higher-dimensional generalization of the Hamiltonian~(\ref{eq:H_NP}), where the random couplings $J_{ij}$ factorize, is known as the Mattis glass, see e.g. Ref.~\cite{Gurarie:2004}. 

For the present problem, we complement the algorithm outlined in section~\ref{subsec:mf_AOA} with the following strategy: Given the solution string $\boldsymbol{\sigma}_*$, try whether flipping any two spins $(n_i^z, n_j^z)$, $i=1, ..., N-1$, $i < j \leq N-1$ produces a lower energy. Here, we have already taken advantage of the $\mathcal{Z}_2$ symmetry of the problem. Therefore, this strategy produces an additional cost of $(N-1)(N-2)/2$, which leaves the asymptotic scaling of our algorithm unaltered. We find that the string following from the mean-field AOA is usually such that this strategy produces an improved solution. Note that we do not find a similar improvement for the SK model discussed in section~\ref{subsubsec:SK_model}. 

We now analyze the distribution of the approximate costs over $10^4$ realizations. The variable of interest is $c=\sqrt{E_*}$ since the square root of the energy here corresponds to the value of the cost function $C(\mathcal{S})$. From the above-mentioned scaling $\propto 2^{-N}$ of the minimal cost of the partition problem, it also follows that it is possible to assume $E_0=0$ as the true minimum. Including the post-evolution strategy of flipping two spins, we thus find an exponential distribution
\begin{align}
    P(c) = (N^\omega/A) \exp{\left(-c N^\omega/A\right)},
\end{align}
where the values of the parameters are $A=3.1$ and $\omega=1.9$ as given in Fig.~\ref{fig:NP_exp_distro}. The first moment of this distribution, plotted in the upper panel of Fig.~\ref{fig:NP_exp_distro}, is $\left\langle \sqrt{E_*}\right\rangle = AN^{-\omega}$. This can now be compared with the results following from the QAOA, shown in Fig.~\ref{fig:NP_QAOA} for different numbers of layers. Compared to its mean-field counterpart, if $p$ is finite, the QAOA shows a much worse scaling $N^{\omega'}$, with a positive scaling exponent $\omega' \approx 1/3$.

In close analogy to the analysis done for the SK model, one can estimate the accuracy of our new classical algorithm for the partition problem. The failure probability $P_f(c)$ to find the optimum above the threshold $c$ is exponential, cf. Eq.~(\ref{eq:P_f_SK}),
\begin{equation}
    P_f(c) = \int_c^{+\infty} dc' P(c') = \exp{\left(-c N^\omega/A\right)}.
\end{equation}
By picking $c=N^{-\delta} \ll 1$  with $\delta = \omega/2$, we arrive at the following conservative estimate:
\begin{enumerate}
\item[]
{\it
With a probability of at least $1 - \exp{(- N^{\delta}/A)}$ over possible realizations of random sets ${\cal S}$ in the partition problem, the mean-field AOA delivers an approximate optimum $c\leq N^{-\delta}$, where the exponent $\delta = 0.95$.
}
\end{enumerate}
In passing, we note that the heuristic classical algorithm due to Karmarkar and Karp~\cite{Karmarker:CSD-83-113, Yakir:1996} performs slightly better --- 
it finds an optimum with an accuracy of $N^{- \Theta\, \ln N}$, where $\Theta$ is a numerical constant of order unity.

To summarize, we have found that the mean-field AOA is capable of identifying optimization problems for which the QAOA should not be expected to yield quantum advantage. Also, at least for the optimization problems investigated here, our results suggest that larger values of $N$ typically make a random instance {easier} to address within the mean-field approach. In other words, in the interesting quantum regime (i.e.\ for system sizes which cannot be simulated), the solution of these optimization problems is unlikely (on average) to acquire an advantage from quantum fluctuations. On one hand, it is known that this does not hold for all cases, i.e.\ there are problems for which the minimal gap is suppressed {exponentially} in $N$; on the other hand, one can ask into which category among these two options most of the real-world problems are likely to fall.

\section{\label{sec:path_int}Path-Integral Approach}

In this section, we go one step beyond the mean-field approximation by studying the Gaussian quantum fluctuations around mean-field spin trajectories. 
The motivation behind this analysis is the hope to delineate `easy' from `hard' instances by looking at the spectrum of the fluctuations, 
thus setting the stage for further exploration of possible quantum advantages.

Path integrals are a well-known tool for describing (quantum) fluctuations around a classical or mean-field trajectory. To derive the path integral for spin degrees of freedom, we employ spin coherent states~\cite{Stone:2000}, as they facilitate the systematic expansion around the mean-field. In this work, 
we limit our analysis to the Gaussian case. This enables us to study the spectrum of `paramagnons` as well as the Lyapunov exponents of the corresponding (one-particle) Green functions.

Even though the mean-field AOA allows for very large $p$ and thus for potentially nearly adiabatic evolution, we expect there to exist `hard' instances for which the gap becomes (exponentially) small, which in turn is likely to render even very slow evolution ultimately non-adiabatic. Our goal in this section is to provide a tool for telling these instances apart from the `easy' ones.

\subsection{Spin Coherent-State Path Integral}

To simplify the discussion, in this section we adopt the standard Hamiltonian of AQC, i.e.\ 
\begin{align}
    \hat H(s) = (1-s) \hat H_{D} + s \hat H_{P},
\end{align}
where now $s=t/T$ and $\hat H_{D, P}$ are defined in Eq.\ \eqref{eq:H}. The total time $T$ of the adiabatic protocol is very long ($T \to \infty$), and the initial ground state is given, as before, by Eq.~\eqref{eq:psi_in}. For the system of $N$ qubits (classical spins), the spin coherent state is defined as the Kronecker product
\begin{align}\begin{split}
\label{eq:g_state}
|{g}\rangle = |g_1\rangle \otimes |g_2\rangle \otimes \cdots \otimes |g_N\rangle,
\end{split}\end{align} 
where the coset element $g_i \in {\rm SU}(2)/{\rm U}(1)$ 
describes the Bloch sphere of the $i$th qubit. The density matrix of the many-qubit system evolves as
\begin{align}\begin{split}
    \hat\rho(t) &= \hat U(t)|{g}_0\rangle \langle {g}_0| \hat U^\dagger(t), \\
    \hat U(t) &= \hat T \exp{\left(-i \int_0^t ds\, \hat H(s)\right)},
\end{split}\end{align}
where $\hat T$ denotes time-ordering and $|{g}_0\rangle$ is the spin coherent-state representation of the initial state $|\psi_0\rangle$ defined in Eq.~\eqref{eq:psi_in}. Note that details on our definition of spin coherent states are provided in Appendix\ \ref{app:spin_path_coherent}.

We now formulate the system evolution via a path integral. In terms of the density matrix, this would require the Schwinger-Keldysh formalism~\cite{Altland:2010, kamenev2023field, Diehl-Keldysh}. Instead, to simplify the discussion, we focus on transition amplitudes $\mathcal{A}(T) = \langle {g}_f| \hat U (T)|{g}_0\rangle$, where $\langle {g}_f|$ is the final spin coherent state. This will prove sufficient for our purposes in the present work. Going over to the path-integral representation in a standard manner~\cite{Altland:2010}, we split the total evolution $U(T)$ into $p \gg 1$ steps, with $\tau = T/p$ being the duration of a single Trotter step, and then use the spin coherent-state resolution of the identity $p$ times.
Upon taking the continuum limit, one then arrives at   
\begin{align}
\label{eq:PathI_A}
\mathcal{A}(T) &= \int_{\boldsymbol{g}_0}^{\boldsymbol{g}_f} \mathcal{D}g\; \exp\left\{ -(S_I + S_H) \right\} \\
&= \int_{\boldsymbol{g}_0}^{\boldsymbol{g}_f} \mathcal{D}g\; \exp\left\{ i \int_0^T dt \langle {g}| [i\partial_t  - \hat H(t)]|{g}\rangle  \right\}, \nonumber
\end{align}   
where $\mathcal{D}g$ is a functional integration measure over all spins and time slices, and constructed following either Eq.~\eqref{eq:I_res_angles} or \eqref{eq:I_res_z}.

The first term of the action in Eq.~\eqref{eq:PathI_A}, $S_I$ is the  {Berry phase}, for which we provide several representations in Appendix \ref{app:spin_path_berry}. When expressed in terms of the Bloch vectors $\boldsymbol{n}_{i}(t)$ (cf.\ Appendix \ref{app:spin_path_H}), the Hamiltonian part of the action becomes
\begin{align}\begin{split}
\label{eq:S_H_n_s}
S_H &= -{is}  \sum_{i=1}^N \int_0^T dt\; \bigg[ h_i + \sum_{j>i}  J_{ij}   n_j^z(t) \bigg] n_i^z(t) \\
&- i (1-s)  \sum_{i=1}^N \int_0^T dt\; \Delta_i n_i^x(t) . \\
\end{split}\end{align}

An interesting remark is in order here: The classical Larmor equations can be derived by imposing the ${\rm SU}(2)$-like Poisson bracket on the Bloch vectors.
Namely, let us consider the Hamiltonian part of the action,
\begin{align}\begin{split}
S_H = i \int_0^T dt H(\boldsymbol{n},\, s), 
\end{split}\end{align}
where $H$ is expressed solely through the Bloch vectors of individual spins, see Eq.~(\ref{eq:S_H_n_s}), and let us  {define} a Poisson bracket as
\begin{align}\begin{split}
\label{eq:SU(2)_Poisson}
\frac 12 \left\{n^\alpha_{i}, n^\beta_{j} \right\} = \delta^{ij} \epsilon_{\alpha\beta\gamma} n^\gamma_{i}.
\end{split}\end{align}
Then the Larmor equations of motion follow from the Hamiltonian principle
\begin{align}\begin{split}
\label{eq:dt_n}
\partial_t n^\alpha_{i} = \left\{n^\alpha_{i},H(\boldsymbol{n}, s)\right\},
\end{split}\end{align}
where the greek indices run through $x,y,z$. The role of the Berry phase $S_I$ is therefore to generate the Poisson bracket~(\ref{eq:SU(2)_Poisson}) when the variational principle is applied to the full action, $\delta S= \delta S_I + \delta S_H$. The saddle-point trajectories of the action thus obey the equations of motion~\eqref{eq:eom_mf} with $\beta\equiv(1-s)$ and $\gamma\equiv s$.

\subsection{Fluctuations Around Mean-Field}

In this subsection, we derive the action of Gaussian fluctuations around the mean-field trajectories. We then use it to estimate how the fluctuations grow in time, and show that the latter can be used an effective tool to differentiate between `hard' and `easy' instances of an optimization problem. Finally, we demonstrate this in some detail for the SK model.

\subsubsection{The action of Gaussian fluctuations}

How can spin quantum fluctuations be parameterized from a geometrical perspective in the most efficient way? To answer this question, as detailed in Appendix~\ref{app:spin_path_stereo}, we employ the stereographic projection~(\ref{eq:z_via_n}) of each spin's Bloch sphere onto the complex plane, thus introducing complex coordinates $(z_i(t), \bar z_i(t))$, which in turn generate the coset elements $\hat g_i(t)$ for each spin via 
\begin{align}\begin{split}\label{eq:g_def_z}
\hat g_i &=  \frac{1}{(1+|z|^2)^{1/2}}\left(
\begin{array}{cc}
1 & -\bar z_i \\
z_i  & 1
\end{array}
\right).
\end{split}\end{align} 
We assume that the saddle-point trajectories $\boldsymbol{n}_i(t)$ of all spins are known to us by virtue of Eq.~(\ref{eq:n(t)}).

Quantum fluctuations in the path integral are due to trajectories $\hat g_i'(t)$ that are close to $\hat g_i(t)$. We hence introduce a  {shifted} coset element as
\begin{align}
\label{eq:fluct_def}
\hat g_i' \hat\sigma^z (\hat{g}_i')^{-1} = \hat g_i \tilde g_i \hat\sigma^z (\hat g_i \tilde g_i )^{-1}, 
\end{align} 
where $\tilde g_i$ is close to the north pole,
\begin{align}\label{eq:g_tilde}
\tilde g_i =  \frac{1}{(1+|\eta_i|^2)^{1/2}}\left(
\begin{array}{cc}
1 & -\bar\eta_i \\
\eta_i  & 1
\end{array}
\right), \quad |\eta_i| \ll 1.
\end{align}
Pictorially, the trajectory $\hat g_i'$ is thus displaced from $\hat g_i$ similarly as $\eta_i$ is displaced from the north pole. Mathematically, the relation~(\ref{eq:fluct_def}) means that $\hat g_i' \sim \hat g_i \tilde g_i$, where the equivalence is understood in the sense of the coset structure, i.e.\ up to right multiplication by any $\hat h$ commuting with $\hat\sigma^z$ (if $\hat g_1 \sim \hat g_2$ then $\hat g_1 = \hat g_2 \hat h$ with $\hat h \hat\sigma^z = \hat\sigma^z \hat h$). The coordinates $(\eta_i, \bar \eta_i)$ are used in the following to parameterize the Gaussian fluctuations around the mean-field solutions.

Assume further that $\hat g_i'$ is expressed via complex coordinates $(z_i', \bar z_i')$. Comparison of Eq.~\eqref{eq:fluct_def} with Eq.~\eqref{eq:g_def_z} then gives
\begin{align}
z_i' = \frac{z_i + \eta_i}{1- \bar z_i \eta_i}, \quad \bar z_i' = \frac{\bar z_i + \bar \eta_i}{1- z_i \bar \eta_i}.
\end{align}
These identities establish the complex coordinates of the shifted trajectories in terms of the coordinates of the original ones, while the fluctuations are parameterized by $\eta_i$. When the latter are small, one expands
\begin{align}
\label{eq:delta_zi}
z'_i = z_i + \delta z_i = z_i + (1+|z_i|^2)(\eta_i + \bar z_i \eta_i^2) + \mathcal{O}(\eta_i^3).
\end{align} 
The relation between $\delta z_i$ and $\eta_i$ is hence non-linear, the rationale behind this being that the path-integral measure is preserved, provided one goes from integration over $z_i'$ to $\eta_i$ at fixed saddle-point trajectory. Furthermore, we note that in the Gaussian regime ($|\eta_i| \ll 1$), the new measure in the variables $\eta_i$ becomes flat, i.e.\
\begin{align}
    \int {dz_i'd\bar z_i'}{\left[1 + \left|z_i'\right|^2\right]^{-2}}  \longrightarrow \int d\eta_i d\bar \eta_i.
\end{align}
To discuss the fluctuation, we introduce the action $S$ in complex representation as
\begin{align}\begin{split}
\label{eq:S_IH_z}
S &= S_I + S_H \\
&= \frac 12 \sum_i \int_0^T dt\,\frac{\dot z_i \bar z_i - z_i {\dot{\bar{z_i}}}}{1 + |z_i|^2}  + i\int_0^T dt\, H(z,\bar z),
\end{split}\end{align}  
where $H(z,\bar z)$ is the complex representation of the Hamiltonian from Eq.~\eqref{eq:S_H_n_s}, which can be calculated by utilizing Eqs.~\eqref{eq:n_via_z}. As shown in Ref.~\cite{Stone:2000}, the classical path, which follows from extremization of the action, obeys the following Hamiltonian equations:
\begin{align}
\label{eq:eq_of_motion_z} 
\dot{z_i} = - i\left(1+|z_i|^2\right)^2 \frac{\partial H}{\partial \bar z_i}, \quad {\dot{\bar{z_i}}} = i \left(1+|z_i|^2\right)^2 \frac{\partial H}{\partial z_i},
\end{align}
which is an equivalent representation of the mean-field equations~(\ref{eq:eom_mf}).
To derive the action of the Gaussian fluctuations ${\cal S}[{\eta}, {\bar\eta}]$ around
this classical path, one parameterize the variations $\delta z_i$ in terms of the $\eta_i$ as derived above in Eq.~\eqref{eq:delta_zi}. 
The calculations detailed in Appendix~\ref{subsec:A_flucs} yield 
the following result:
\begin{align}
\label{eq:S_fluctuations}
{\cal S}[{\eta}, {\bar\eta}] = \frac i 2 \int_0^T dt\, ({\bar\eta}\; {\eta}) 
\left[
\begin{array}{cc}
- i\partial_t + A & B \\
B^\dagger & i\partial_t + \bar A
\end{array}
\right] 
\left(
\begin{array}{c} {\eta} \\ {\bar\eta} \end{array}
\right),
\end{align}
where the matrices $A(t)$ and $B(t)$ are time-dependent through their dependence on the classical path,
and $\eta$ and $\bar \eta$ are $N$-dimensional spinors constructed from $\eta_i$ and $\bar\eta_i$, respectively. An important comment is in order here: When analyzing the dynamics of the fluctuations, we found that it is crucial to parameterize the mean-field trajectories such that the final mean-field AOA solutions $(\boldsymbol{\sigma_*})_i = \operatorname{sign}(n_i^z(T))$ are stereographically projected onto the  {origin} of the complex plane. In other words, the poles $p_i$ of the Bloch sphere from which to perform the projection for each spin trajectory are defined as $p_i := (0, 0, -\operatorname{sign}(n_i^z(T))$. Under this convention, after again substituting `Cartesian' coordinates on the sphere, we find for the above matrices the following components:
\begin{align}
\label{eq:AB_diag}
A_{ii} =  \frac{2 (1-s)\Delta_i n_i^x}{1 + (\boldsymbol{\sigma_*})_i^{} n_i^z} + 2 s (\boldsymbol{\sigma_*})_i^{} m_i, \qquad
B_{ii} = 0,
\end{align}
where again $s=t/T$ and the self-consistent magnetization $m_i$ was defined in Eq.~\eqref{eq:magnetization}. With the shorthand notation $ n_i^{\pm}= (\boldsymbol{\sigma_*})_i^{} n_i^x \pm i n_i^y$, one finds for the off-diagonal components that
\begin{align}\begin{split}
\label{eq:AB_off-diag}
A_{ij} = -s\, J_{ij} n_i^+ n_j^-, \quad B_{ij} = -s\, J_{ij} n_i^+ n_j^+,
\end{split}\end{align}
such that ${A} = A^\dagger$ is Hermitian and $ B =  B^T$ is symmetric. Hence, the effective Hamiltonian becomes
\begin{align}
\label{eq:H_t}
\hat{\mathcal{H}}(t) = \left(
\begin{array}{cc}
 A(t) &  B(t) \\ 
 B^\dagger(t) & \bar{A}(t)
\end{array}
\right), \quad
\hat\tau_3 = \left(
\begin{array}{cc}
\mathds{1} & \mathds{0}  \\  & -\mathds{1}
\end{array}
\right), 
\end{align}
where we have also introduced the matrix $\hat\tau_3$ acting on the spinor space of $(\boldsymbol{\bar\eta}, \boldsymbol{\eta})^T$ defining the block decomposition of $\hat{\mathcal{H}}$.

The instantaneous spectrum of `paramagnons', $\omega_\mu(s)$, $\mu=0, ..., 2N-1$, at given time $s=t/T$ 
can be then found via the (positive) eigenvalues of the operator $\hat\tau_3 \hat{\mathcal{H}}$, namely
\begin{align}
\label{eq:Omega_s}
{\rm det}\left[\omega_\mu(s)- \hat\tau_3 \hat{\mathcal{H}}(s)\right] = 0.
\end{align}
The smallest eigenvalue, $\omega_0(s)$, may serve as an indicator of the gap between the ground and the first excited state of the many-body Hamiltonian $\hat H(s) = (1-s) \hat H_{D} + s \hat H_{P}$.

\subsubsection{The dynamics of quantum fluctuations}

The magnon spectrum~(\ref{eq:Omega_s}) can only determine the stability of the instantaneous ground state and thus fails in the non-adiabatic regime. The latter is, however, precisely the point of interest when the mean-field AOA does not perform well. To address this issue, we resort to the equation of motion for the {Green function}, which we define as
\begin{align}
\boldsymbol{G}(t, t') := -i\left\langle {\boldsymbol{\eta}}(t)^{} \otimes {\boldsymbol{\eta}}^\dagger(t') \right\rangle,
\end{align}
where ${\boldsymbol{\eta}}^T = (\eta, \bar\eta)$ is the $2N$-component spinor and the average is done with respect to the Gaussian action Eq.~\eqref{eq:S_fluctuations}. The equation of motion for the Green function can be written in two complementary forms,
\begin{align}
\label{eq:G_s_1}
        \left[i\hat\tau_3\overrightarrow{\partial_t} - \hat{\mathcal{H}}(t)\right]\boldsymbol{G}(t, t') &= \mathds{1}\delta(t - t'), \\
\label{eq:G_s_2}
        \boldsymbol{G}(t, t')\left[-i\hat\tau_3\overleftarrow{\partial_{t'}} - \hat{\mathcal{H}}(t')\right] &= \mathds{1}\delta(t - t').
\end{align}
To specify $\boldsymbol{G}(t, t')$ uniquely, the system of these differential equations needs to be supplemented by appropriate 
boundary conditions at $t=0$ and $t=T$. The latter have to be treated carefully since we are dealing with first-order rather than second-order differential operators. As discussed in Ref.~\cite{Stone:2000}, the boundary conditions for the fluctuations assume the form $\eta_i(0) = \bar \eta_i(T) = 0$, while $\bar\eta_i(+0)$ and $\eta_i(T-0)$ are, in fact, unbounded independent integration variables. Expressed in vector form, they translate into
\begin{equation}
(\mathds{1} + \hat\tau_3) \boldsymbol{\eta}(0) = 0, \qquad  \boldsymbol{\eta}^\dagger(T) (\mathds{1} + \hat\tau_3)  = 0,
\end{equation}
and, when applied to the Green function, they become
\begin{equation}
\label{eq:bc_G}
(\mathds{1} + \hat\tau_3) \boldsymbol{G}(0, t') = 0, \qquad 
\boldsymbol{G}(t, T) (\mathds{1} + \hat\tau_3) = 0,
\end{equation}
where $t'>0$ and $t<T$. 

With these preliminaries at hand, we are now in position to write down a formal solution to Eqs. (\ref{eq:G_s_1}-\ref{eq:G_s_2}). 
To this end, note that the Green function is discontinuous at equal times $t=t'$, with the jump 
\begin{equation}
i\hat\tau_3 (\boldsymbol{G}(t+0, t)-\boldsymbol{G}(t-0, t)) = \mathds{1}.
\end{equation}
Hence for $t' \to t$, we can write
\begin{equation}
\label{eq:G_eq}
\boldsymbol{G}(t, t') = {\boldsymbol{{g}}}(t, t') - \frac{i}{2} \hat\tau_3 \, {\rm sgn}(t-t'),
\end{equation}
where $\boldsymbol{{g}}(t, t')$ is the continuous part of the Green function. This ansatz enables us to introduce
the correlator $\boldsymbol{g}(t)$ at coinciding time points defined by the relation
\begin{equation}
\boldsymbol{g}(t) = 2 i \lim_{t' \to t} \boldsymbol{g} (t,t') \, \hat\tau_3.
\end{equation}
One can prove that this correlator fulfils the normalization constrain, $\boldsymbol{g}^2(t) = \mathds{1}$, and 
satisfies the much simpler differential equation
\begin{align}\label{eq:equal_time_F}
    \begin{split}
        i\partial_{t}\boldsymbol{g}(t) &= \left[\hat{\mathcal{L}}(t),  \boldsymbol{g}(t) \right], 
        \quad {\cal\hat L}(t):= \hat\tau_3 {\cal\hat H}(t).
    \end{split}
\end{align}
To derive the above result, one subtracts Eq.~\eqref{eq:G_s_1} from \eqref{eq:G_s_2} and takes the equal-time limit $t' \to t$.
This is a first-order differential equation, which as before requires some boundary conditions. The latter can be inherited from the ones stated in Eq.~(\ref{eq:bc_G}). By setting $t'=+0$ and $t=T-0$, one reduces them to
\begin{align}\label{eq:bc_Q_matrix}
    \begin{split}
        (\mathds{1} + \hat\tau_3) \left(\boldsymbol{g}(0) - \mathds{1}  \right) &= 0, \\
        \left(\boldsymbol{g}(T) - \mathds{1} \right) (\mathds{1} + \hat\tau_3) &= 0.        
    \end{split}
\end{align}
It is worth mentioning here that if the time $t$ is substituted by a spatial coordinate $x$, then Eq.~(\ref{eq:equal_time_F}) 
turns into the quasiclassical Eilenberger equation in the theory of superconductivity~\cite{Shelankov:1985}. 
Specifically, $N$ plays the role of the number of transport channels in a quasi-one-dimensional geometry, which is relevant for studies
of the Josephson's effect across superconducting weak links or point-like junctions, while the two-component structure of the spinor $\boldsymbol{\eta} = (\eta, \bar \eta)^T$ is analogous to the 
decomposition of an electron wave function into left- and right-traveling wave packets with momenta lying close to the Fermi surface.
In this picture, the matrices $ A$ and $ B$ describe, respectively, forward and backward inter-channel scattering due to disorder. Besides, the form of the boundary conditions~(\ref{eq:bc_Q_matrix}) exactly 
matches the ones imposed on the Green function within the quasi-classical framework~\cite{Nazarov:1999, Neven:2013}. 

With these remarks in mind, we proceed by solving Eq.~(\ref{eq:equal_time_F}) using the scattering formalism of Ref.~\cite{Beenakker:1997}.
For that, we introduce a time-dependent scattering matrix $\boldsymbol{M}(t)$ which by definition satisfies the equation
\begin{equation}
i\partial_{t}\boldsymbol{M}(t) = \hat{\mathcal{L}}(t) \boldsymbol{M}(t), \qquad \boldsymbol{M}(0)=\mathds{1},
\end{equation}
which is formally solved by the time-ordered exponential
\begin{equation}
\boldsymbol{M}(t) = \hat T \, e^{ - i \int_0^t d\tau \hat{\mathcal{L}}(\tau)}.
\end{equation}

In our numerical implementation of the algorithm, one can effectively
find $\boldsymbol{M}(t)$ by means of Trotterization,
\begin{equation}
\boldsymbol{M}(t_k) = \prod_{j=1}^k e^{ - i \hat{\mathcal{L}}(t_j) \tau }, \qquad t_k = k \tau,
\end{equation}
with $\tau = T/p$ and $p \gg 1$ as before. Now Eq.~\eqref{eq:equal_time_F} is solved by 
\begin{align}
    \boldsymbol{g}(t) = \boldsymbol{M}(t)\boldsymbol{g}(0)\boldsymbol{M}(t)^{-1},
\end{align}
and the transfer matrix in its canonical form~\cite{Beenakker:1997} can be shown to be diagonalizable as
\begin{align}
\label{eq:MM_eigenbasis}
    \boldsymbol{M}(t)\boldsymbol{M}(t)^\dagger &= \boldsymbol{U}\operatorname{diag}\left(e^{-2\boldsymbol{\lambda}(t)},\; e^{2\boldsymbol{\lambda}(t)}\right) \boldsymbol{U}^\dagger,
\end{align}
where $\boldsymbol{U}$ is a unitary matrix and 
\begin{align}
    \boldsymbol{\lambda}(t) = \left(\lambda_0(t), ..., \lambda_{N-1}(t)\right)^T
\end{align} 
are the set of positive Lyapunov exponents we look for. We also note that $\boldsymbol{M}$ obeys to the `flux-conservation condition' 
\begin{align}\label{eq:flux_conservation}
    \boldsymbol{M}(t)^\dagger \hat\tau_3 \boldsymbol{M}(t) = \hat\tau_3, 
\end{align}
which stems from the Hermiticity of the underlying Hamiltonian. 

With the transfer matrix at hand, one can write $\boldsymbol{g}(T) = \boldsymbol{M}(T) \boldsymbol{g}(0) \boldsymbol{M}^{-1}(T)$ and further use this relation together with boundary conditions~(\ref{eq:bc_Q_matrix}) to find the {unknown} $\boldsymbol{g}(0)$ and $\boldsymbol{g}(T)$. The general solution is rather involved, see Refs.~\cite{Nazarov:1999, Neven:2013} for more details. However, for many instances of the SK model which we study below, the Lyapunov exponents satisfy $\lambda_l(0) = \lambda_l(T) = 0$ for all $l$, which is the hallmark of reflectionless scattering (for more details, see Appendix~\ref{sec:M_matrix}). Under this condition, the above set of equations is solved by $\boldsymbol{g}(0) = \boldsymbol{g}(T) = \hat\tau_3$, which we adopt in what follows.

\begin{figure}[t!]
	\begin{center}
		\includegraphics[width=.45\textwidth]{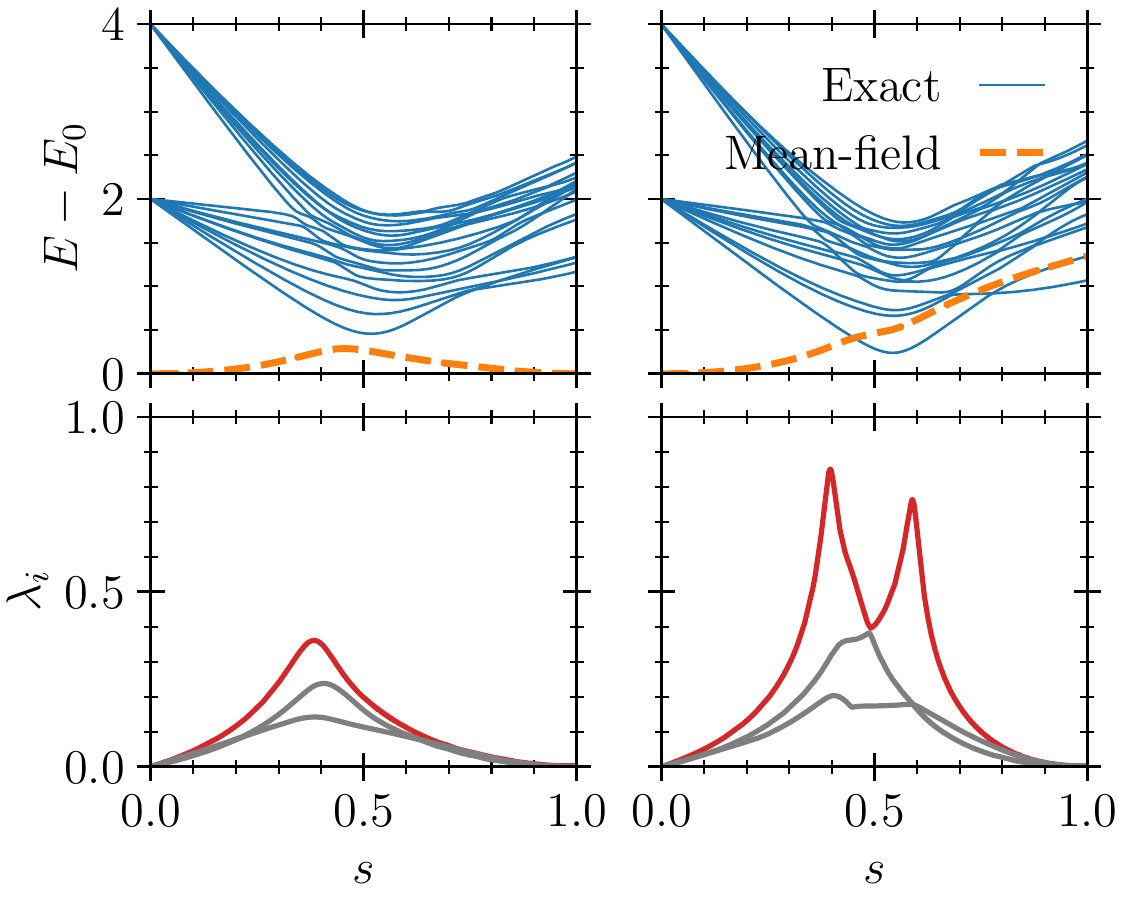}
		\caption
		{\label{fig:easy_hard} The exact eigenspectrum (upper panels) vs.\ the Lyapunov exponents (lower panels) for two concrete instances of the SK model at $N=11$. For the `easy' instance (left panels), the schedule parameters are $\tau=1/2$, $p=2\cdot 10^3$; for the `hard' instance (right panels),  we have instead $\tau=1/2$, $p=5\cdot 10^3$. The dashed lines in the upper panels show the energy $E_*$ returned by the mean-field AOA. The threshold in Eq.~\eqref{eq:largest_lyapunov} evaluates to $\lambda_0(t_*) \sim 1.2$. Only the three largest Lyapunov exponents are shown.}
	\end{center}
\end{figure}

\begin{figure}[t!]
	\begin{center}
		\includegraphics[width=.45\textwidth]{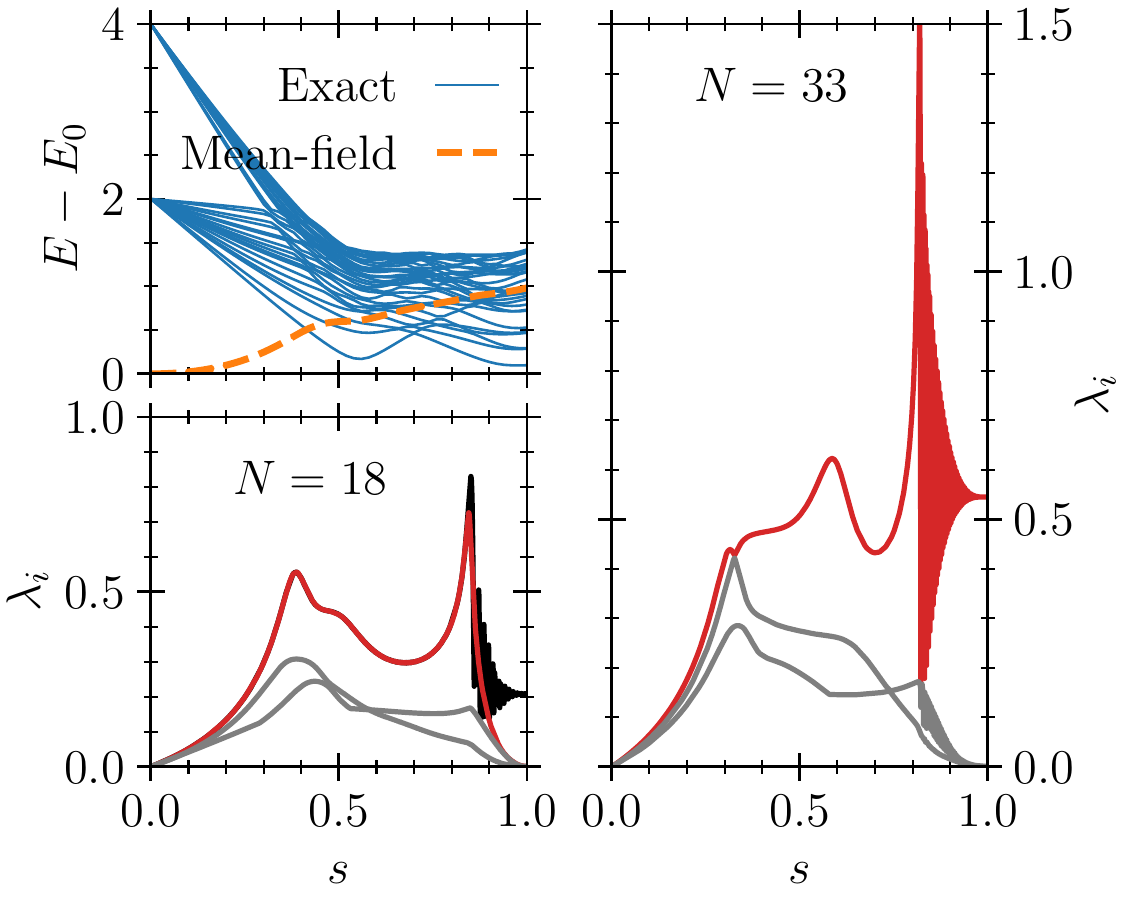}
		\caption
		{\label{fig:hard_large} The exact eigenspectrum (upper left panel) versus the Lyapunov exponents (lower left and right panel) for two concrete instances of the SK model at $N=18$ and $N=33$. For $N=18$ (left panels), the schedule parameters are $\tau=0.5$, $p=2\cdot 10^4$; for $N=33$ (right panel), we have instead $\tau=0.4$, $p=2\cdot 10^4$. The dashed line in the upper left panel shows the energy $E_*$ returned by the mean-field AOA, while the oscillatory (black) line in the lower left panel shows the maximum exponent for $\tau=0.5$, $p=5\cdot 10^3$.}
	\end{center}
\end{figure}

We are now in position to estimate the quantum fluctuations above the mean-field solution. The simplest quantity to assess 
is~\footnote{We regularize the equal-time average as $\langle|\eta_i(t)|^2\rangle \equiv \frac 1 2 \langle\eta_i(t) \bar\eta_i(t+0) + \bar\eta_i(t) \eta_i(t+0)\rangle$.}
\begin{align}
        \sum_{i=1}^N \langle|\eta_i(t)|^2 \rangle = \frac 12 {\operatorname{Tr} \boldsymbol{g}(t) \hat\tau_3}
        = \frac 1 2 \!\operatorname{Tr} \!\left(\boldsymbol{M}(t)\boldsymbol{g}(0) \boldsymbol{M}^{-1}(t) \hat\tau_3\right).
\end{align}
On substituting $\boldsymbol{g}(0) = \hat\tau_3$ and with the use of relation~(\ref{eq:flux_conservation}), one finds
\begin{align}
        \sum_{i=1}^N \langle|\eta_i(t)|^2\rangle = \frac 12 \operatorname{Tr} \,\boldsymbol{M}(t) \boldsymbol{M}^\dagger(t)
        = \sum_{l=0}^{N-1} \cosh 2\lambda_l(t).
\end{align}
At this point, we rely on empirical evidence suggesting that when quantum fluctuations grow in time (cf.\ Figs.~\ref{fig:easy_hard}, \ref{fig:hard_large}), the sum above is dominated by the maximal Lyapunov exponent $\lambda_0(t)$, which is supported by our numerical analysis. Furthermore, in a disordered system with strong graph connectivity, all correlations are expected to be site-independent when considered by order of magnitude, allowing us to estimate
\begin{equation}
\langle|\eta_i(t)|^2\rangle \sim \frac 1 N \, e^{2 \lambda_0(t)}.
\end{equation}
As one can see, the mean-field approximation works well provided $\lambda_0(t) \lesssim \tfrac 12$. In this regime, fluctuations are suppressed by a factor $1/N$, the latter parameter thus effectively playing a role of $\hbar$ in our semi-classical approximation to the QAOA. However, the mean-field AOA entirely breaks down if at a certain time $t_*$ quantum fluctuation become sizable, i.e.\ $\langle|\eta_i(t_*)|^2\rangle \sim 1$. This happens when the largest Lyapunov exponent reaches the value
\begin{equation}\label{eq:largest_lyapunov}
\lambda_0(t_*) \sim \ln \sqrt{N}.
\end{equation}

We now investigate the properties of the Lyapunov exponents $\boldsymbol{\lambda}$ for several instances of the SK model. As shown in Fig.~\ref{fig:easy_hard} for $N=11$, it is indeed possible to differentiate between `easy' and `hard' instances on the basis of these exponents. For the former case (left panels of Fig.~\ref{fig:easy_hard}), the mean-field AOA is found to return the exact ground state, while the $\boldsymbol{\lambda}$ remain small. Even here, however, the shrinking of the gap is accompanied by an increase of the Lyapunov exponents. For the `hard' instance (right panels of Fig.~\ref{fig:easy_hard}), the mean-field AOA wrongly returns the second excited state as a solution. In this case, both the first crossing of the exact levels and the closing of the gap are accompanied by a sharp increase in the Lyapunov exponents. The estimate from Eq.~\eqref{eq:largest_lyapunov} shows that both maxima of 
the largest Lyapunov exponent for this instance are just slightly below threshold and thus the spin system finds itself in a regime of strong quantum fluctuations. 

Before discussing our simulations for larger system sizes (Fig.~\ref{fig:hard_large}), the following comments are in order. 
The first mini-gap in the exact spectrum of the adiabatic Hamiltonian $\hat H(s)$ (in the case of the SK model 
it is located approximately at $s_* \simeq 0.5$ as seen from Fig.~\ref{fig:easy_hard}) 
is the hallmark of the ergodic-to-MBL quantum phase transition between a delocalized paramagnet and a localized spin-glass phase~\cite{Wang2022}. 
This gap is believed to have only a polynomial scaling with respect to the system size, $1/N^\alpha$, where $\alpha>0$ is a critical exponent
\footnote{
For the closely related fully connected Hopfield model, it was argued recently~\cite{Knysh2016} that $\alpha=1/3$.
}.
For larger random instances, it is understood that \textit{subsequent} small-gap bottlenecks appear deep in the MBL-phase close to the end of the adiabatic algorithm~\cite{Altshuler2010}. As opposed to the first mini-gap, they are exponentially small in $N$ for NP-hard combinatorial optimization problems.
For the Hopfield model, which is a close analog of the SK model, such (stretched) exponential laws in $N$ have also been conjectured in Ref.~\cite{Knysh2016}.

The appearance of the third sharp peak in the largest Lyapunov exponent $\lambda_0(s)$ (see Fig.~\ref{fig:hard_large}) is a semi-classical counterpart of the above scenario related to the fact that $\omega_0(s_0)$ may become exponentially small in $N$ at some point $s=s_0$, indicating the presence of a `hard' instance. The adiabaticity condition of the mean-field AOA in this case is broken provided that the run time, $T= p \tau$, is not sufficiently long, $T < {2 \pi}/{\omega_0(s_0)}$. Under this condition, $\lambda_0(s)$ develops time-dependent oscillations at $s>s_0$, 
which can be removed by choosing a longer $T$ to restore adiabaticity. However, the peak as such remains present and a noticeable improvement of the approximate optimum $E_*$ is not guaranteed.
The sharp extrema in $\lambda_0(s)$ become progressively larger for `hard' instances as $N$ increases, although, as we have found, the logarithmic threshold~(\ref{eq:largest_lyapunov}) grows accordingly and is never violated.

The left panels of Fig.~\ref{fig:hard_large} illustrate the outlined story for $N = 18$ in concrete terms. In agreement with the sharp spike of the largest Lyapunov exponent toward the end, the mean-field AOA does not return the correct ground state. The main feature of Fig.~\ref{fig:hard_large} is, however, the characteristic oscillations of the largest eigenvalue for $p = 5\cdot 10^3$ (black line in the background). As highlighted by the smooth (red) line in the foreground, these oscillations disappear for a very large value of $p = 2\cdot 10^4$. One should stress that these oscillations {\it do not} accompany the transition of the system to a chaotic regime. However, in this case the algorithm returns a poor approximate optimum described by the rare-event statistics~(\ref{eq:P_f_SK}). 
In the case of an even larger system size $N = 33$, we find that such oscillations persist even at very large values of~$p$.

\subsection{Adiabaticity Condition}

We close this section by analyzing the complexity of our classical algorithm for large $N$. First of all, note that at times $s>s_*$, i.e.~above the critical point of the ergodic-to-MBL phase transition in the Hamiltonian $H(s)$ of the {\it quantum} spin system, the mobility edge in its many-body spectrum emerges~\cite{Alet:2018, Filho:2022}. 
The states below the mobility edge are many-body localized, while those above are ergodic, which can be diagnosed via their level spacing statistics~\cite{Oganesyan:2007}. For times $s$ approaching the critical point from above, $s \to s_* +0^+$, the mobility edge merges with
the instantaneous ground state $E_0(s_*)$. We can then invoke the classical to quantum correspondence~\cite{Boerner:2023} to identify the ergodic part of the quantum spectrum with the classical chaotic regime of the mean-field spin Hamiltonian~(\ref{eq:H_classical}), and analogously for the complementary MBL part and the classical regular regime. 

These considerations put a certain restriction on the adiabaticity condition of the mean-field AOA. Specifically, the run time $T$ should be long enough,
$T \gg 2\pi/\omega_0(s_*)$, where $\omega_0(s)$ is the smallest eigenvalue in the paramagnon spetrum, see Eq.~(\ref{eq:Omega_s}).
Though it does not happen in the SK model for $N$ of order $10^2$ which we have studied here, the violation of this adiabaticity condition may trigger the crossover of the classical spin dynamics into the chaotic regime at $s>s_*$, accompanied by a sharp increase of the largest Lyapunov exponent $\lambda_0(s)$ that is incompatible with the bound~(\ref{eq:largest_lyapunov}) justifying the semi-classical approximation. Assuming that at $s \to s_*$ the mini-gap
in the paramagnon spectrum behaves as $\omega_0(s_*) \sim 1/N^{\alpha}$, with $\alpha$ being a critical exponent associated with the ergodic-to-MBL phase transition, one concludes that at fixed $\tau$ the number of steps $p$ 
should scale at least as $p \sim N^{\alpha}$. Referring to our previous estimate from section~\ref{subsec:mf_AOA}, we then arrive
at the polynomial complexity of the mean-field AOA,  $O(p N^2 ) = O(N^{2 + \alpha})$. The main conclusion here is that since our classical algorithm
delivers only an approximate optimum of the NP-hard problem (for exact statements see Secs.~\ref{subsubsec:SK_model} and \ref{subsubsec:number_part}), its run time does not scale exponentially in $N$.

\section{\label{sec:discussion_outlook}Discussion \& Outlook}

In this work, we presented a quantum-inspired classical algorithm: When applied to the alternating layers of problem and driver Hamiltonians characteristic of the QAOA, the mean-field approximation yields closed, classical equations of motion that can be solved exactly for any number of layers $p$ and system sizes $N$. Therefore, in contrast to its quantum analog, the mean-field AOA is not limited to very small values of $p$, making it convenient to mimic an annealing-like schedule instead of optimizing over the parameters, as would be the case in the standard QAOA. 

A comparison of the mean-field AOA and the QAOA revealed that the new algorithm can indeed serve as a useful tool to identify optimization problems for which the application of the QAOA could still prove advantageous. That is, for any given problem, if the mean-field AOA does a satisfactory job on finding approximate solutions, little stands to be gained by switching to the full QAOA. A possible strategy for assessing this is to compare the approximate results returned by the two algorithms on exemplary (small) problem instances both among themselves and with available problem-specific classical solvers.

One possible criticism of our results for the SK model is that the mean-field approximation can be expected to perform well given the `self-averaging' properties of the coupling matrix. However, it is not obvious that it should perform  {better} than the QAOA. Furthermore, as mentioned in section~\ref{subsubsec:SK_model}, the SK model was employed only recently~\cite{Farhi:2022} to demonstrate that the QAOA can outperform other classical algorithms at $p=12$ for large $N$. As was demonstrated in Fig.~\ref{fig:Algebraic_decrement_SK}, the mean-field AOA in turn surpasses this benchmark.

Our second benchmark, the partition problem of section~\ref{subsubsec:number_part}, is known to be NP-complete~\cite{mezard2009information}. While it was not in line with the purpose of this work to compare the performance of the new algorithm against other classical algorithms specific to this problem, we showed that the mean-field AOA, supplemented by a spin-flip strategy, gives rise to a well-defined exponential distribution for its output. This scaling works so precisely that one could envision finding an analytical confirmation in future work. The QAOA, in comparison, performs worse on average than the mean-field AOA, even when the additional spin flips are not performed. Given the way the QAOA needs to be implemented on an actual hardware, we also do not expect the spin-flip strategy to improve on the typical bitstrings returned upon measurement. 

In the final part of the paper, section~\ref{sec:path_int}, we went beyond the mean-field approximation and studied the Gaussian quantum fluctuations via a spin coherent-state path integral.  We found very promising results that seem to open up a number of perspectives for follow-up work. Most importantly, we believe the fluctuation analysis can have a useful impact on the schedule design of annealing problems. 

Finally, it could be interesting to pursue the mean-field AOA as a novel optimization algorithm in its own right, e.g.\ by investigating in some detail its performance on the Hopfield model, or by adapting so-called `shortcuts to adiabaticity' to the mean-field framework~\cite{Wang2022}.

\begin{acknowledgments}
The authors acknowledge partial support from the German Federal Ministry of Education and Research, under the funding program "Quantum technologies - from basic research to the market", Contract Numbers 13N15688 (DAQC), 13N15584 (Q(AI)2) and from the German Federal Ministry of Economics and Climate Protection under contract number, 01MQ22001B (Quasim). 
\end{acknowledgments}

\appendix

\section{QAOA vs. Mean-Field AOA}\label{app:QAOA_vs_MF}

This appendix provides further details on the relationship of our algorithm to the standard QAOA. In the latter, one starts from the initial state 
\begin{align}\begin{split}
|\psi_0\rangle = |+\rangle_1^X \otimes |+\rangle_2^X \otimes \cdots \otimes |+\rangle_N^X,
\end{split}\end{align}
where $|\pm\rangle_i^X$ is the $X$-basis of the $i$th qubit.
After the QAOA evolution, the quantum system ends up in the final state, which one decomposes in the $Z$-basis as
\begin{align}\begin{split}
|\psi_f\rangle &= \sum_{\sigma_i =\pm} a(\boldsymbol{\sigma}) |\sigma_1\rangle_1^Z \otimes \cdots \otimes |\sigma_N\rangle_N^Z
\end{split}\end{align}
with complex amplitudes $a(\boldsymbol{\sigma}) = a(\sigma_1, ..., \sigma_N)$.
The sum runs over all of the $2^N$ possible bitstrings. The amplitudes then yield the probabilities
\begin{align}\begin{split}
Q(\boldsymbol{\sigma}) = |a(\boldsymbol{\sigma})|^2, \quad \sum_{\boldsymbol{\sigma}} Q(\boldsymbol{\sigma}) = 1,
\end{split}\end{align}
to measure the system in the respective state.

Now in the mean-field AOA, one instead deals with classical spin vectors
\begin{align}
    {\boldsymbol{n}}^{(i)}(t) = \left(n_i^x(t), n_i^y(t), n_i^z(t)\right)^T
\end{align}
They are normalized to unity, $\left|\boldsymbol{n}^{(i)}(p)\right|^2 = 1$, for all $i=1, ..., N$ and at any time slice $p$ of the algorithm. As mentioned in the main text, the initial condition is ${\boldsymbol{n}}^{(i)}(0) = (1, 0, 0)^T$. The approximate probabilities $P(\boldsymbol{\sigma})$ follow straightforwardly from these vectors. Under the mean-field approximation,  these probablities are factorizable, i.e.\
\begin{align}\begin{split}
P(\boldsymbol{\sigma}) = \prod_{i=1}^N P^{(i)}(\sigma_i), \quad P^{(i)}(\pm) := \frac 12 \left(1 \pm n^z_{i}(p)\right),
\end{split}\end{align}
where again $\sum_{\boldsymbol{\sigma}} P(\boldsymbol{\sigma}) = 1$. Of course, this factorization generally does not hold for the probabilities $Q(\boldsymbol{\sigma})$ extracted from the full quantum approach.

In more detail, in the mean-field framework the $i$th qubit possesses a density matrix
\begin{align}\begin{split}
\hat\rho^{(i)} = \frac 12 \left(
\begin{array}{cc}
1 + n_i^z & n_i^x - i n_i^y \\
n_i^x + i n_i^y & 1 - n_i^z
\end{array}
\right),
\end{split}\end{align}
such that, e.g., for an average $x$-component of the spin one has
\begin{align}\begin{split}
    n_i^x = \operatorname{Tr} \left[\hat\rho^{(i)} \hat\sigma^x_i\right] = \langle  \hat\sigma^x_i \rangle,
\end{split}\end{align}
where it should be understood that the brackets do not signify averaging over problem instances, as in the main text, but the proper quantum average. Similar expressions are valid for the $y$- and $z$-components of the spins. The full density matrix of the system in mean-field approximation also factorizes,
\begin{align}\begin{split}
\hat\rho \equiv \hat\rho^{(1)} \otimes \cdots \otimes \hat\rho^{(N)}.
\end{split}\end{align}
With this ansatz, any spin--spin correlation function over different sites factorizes into a product of averages, i.e.\ its  {irreducible} part is, by definition, missing under the mean-field approximation. This gives rise to another way of expressing this approximation, namely
\begin{align}
    \langle \hat O_i \hat O_j \rangle \equiv \langle \hat O_i \rangle \langle \hat O_j \rangle
\end{align}
for operators $\hat O_i = \hat\sigma^\alpha_{i}$, with $\alpha=x,y,z$, where the brackets again denote the quantum average.

\section{Spin Path Integral}

\subsection{Spin Coherent States}\label{app:spin_path_coherent}

The construction of the spin path integral starts from the introduction of the basis of coherent states. Let the states $|0\rangle$ 
and $|1\rangle$ form the computational basis where the Pauli matrices $\hat\sigma^\alpha$, $\alpha=x,y,z$ are defined in the conventional way.
Then an arbitrary spin coherent state
$|\tilde g\rangle$ can be obtained from $|0\rangle$ by a unitary rotation
\begin{align}\begin{split}
|\tilde  g\rangle = \tilde  g |0\rangle, \quad \tilde  g = e^{- i \phi \hat\sigma^z/2} e^{- i \theta \hat\sigma^y/2} e^{- i \psi \hat\sigma^z/2}, 
\end{split}\end{align}
with the group element $\tilde  g \in \mathrm{SU}(2)$ parameterized in terms of three Euler angles. Note that the role of the angle $\psi$ is merely an extra phase factor, i.e.\
one can always write $|\tilde  g\rangle  = |g\rangle e^{-i\psi/2}$, where 
\begin{align}\begin{split}
\label{eq:g_def}
|g\rangle =  \hat g |0\rangle, \quad  \hat g = e^{- i \phi \hat\sigma^z/2} e^{- i \theta \hat\sigma^y/2}, 
\end{split}\end{align}
and now $g$ is taken from the coset space isomorphic to the two-sphere, $\hat g \in \mathrm{SU(2)/U(1)} \simeq S_2$.
The state $|g\rangle$ is the  {spin coherent state}.
In spherical coordinates $(\theta,\phi)$ it takes the form
\begin{align}\begin{split}
\label{eq:g_sphere}
|g\rangle =|0\rangle \cos\frac \theta 2\, e^{-i \phi/2} + |1\rangle \sin\frac \theta 2\, e^{i \phi/2}.
\end{split}\end{align} 
The collection of these states forms an overcomplete basis, which can be seen from the resolution of identity,
\begin{align}
\label{eq:I_res_angles}
\int_{S_2} \mu(g) |g\rangle \langle g| = \sum_{s = 0, 1}|s\rangle \langle s|, \quad \mu(g) = \frac{1}{2\pi} \sin \theta d\theta d\phi.
\end{align}
Here, by definition, the bra state is
\begin{align}\begin{split}
\langle g| = \langle 0| \hat g^\dagger.
\end{split}\end{align}
Given the state $|g\rangle$, we define the associated density matrix as
\begin{align}\begin{split}
\label{eq:rho_def}
\hat\rho = |g\rangle \langle g| \equiv \frac 12 ( \mathds{1} + \hat q), \quad \hat q: = \hat g \hat\sigma^z \hat g^{-1}.
\end{split}\end{align} 
The matrix $\hat q$ satisfies $\hat q^2 = \mathds{1}$ which in turn implies the purity of the density matrix, $\hat\rho^2 = \hat\rho$, i.e. it is a  {projector}. In spherical coordinates, using (\ref{eq:g_def}), the $\hat q$-matrix becomes
\begin{align}\begin{split}
\label{eq:q_sphere}
\hat q &= \left(
\begin{array}{cc}
\cos\theta & e^{-i\phi} \sin\theta \\
e^{i\phi} \sin\theta & -\cos\theta
\end{array}
\right) \\
&\equiv
\left(
\begin{array}{cc}
n^z & n^x - i n^y \\
n^x + i n^y &  - n^z
\end{array}
\right) \equiv \sum_{\alpha=x, y, z} n^\alpha \hat\sigma^\alpha,
\end{split}\end{align}
where $\boldsymbol{n} = (n^x, n^y, n^z)^T \in S_2$ is the unit vector defining the Bloch sphere. 
 
\subsection{Stereographic Projection}\label{app:spin_path_stereo}
The choice of spherical coordinates to parameterize $g$ is not unique.
One can equivalently define it using complex coordinates $z \in \mathds{C}$,
which we widely use to introduce the quantum fluctuations around the mean-field trajectories in the spin path integral. To this end we define the stereographic projection $\mathcal{P}: \mathds{C}\to S_2$, $z \mapsto (n^x, n^y, n^z)^T  $ from the complex plane to the sphere via
\begin{align}\begin{split}
\label{eq:n_via_z}
n^x \pm i n^y = \frac{2 z}{1 + |z|^2}, \quad n^z = \frac{\pm\left(1-|z|^2\right)}{1 + |z|^2}.
\end{split}\end{align} 
The inverse mapping (from the sphere $S_2$ onto $\mathds{C}$) reads
\begin{align}\begin{split}
\label{eq:z_via_n}
z = \frac{n^x \pm i n^y}{1 \pm n^z} \equiv  \begin{cases} e^{i\phi} \tan \frac\theta 2, \\ e^{-i\phi} \cot \frac\theta 2. \end{cases}
\end{split}\end{align}
Under this mapping, the north pole $(0, 0, 1)$ (south pole $(0, 0, -1)$) is projected onto the origin of $\mathds{C}$, the south pole (north pole) goes to infinity, and the equator becomes the unit circle $|z|=1$. In complex coordinates, one defines the matrices $\hat g$ and $\hat q$ in the following way:
\begin{align}\begin{split}\label{eq:g_def_q}
\hat g &=  \frac{1}{(1+|z|^2)^{1/2}}\left(
\begin{array}{cc}
1 & -\bar z \\
z  & \phantom{-}1
\end{array}
\right),  \\
\hat q &= \hat g \hat\sigma^z \hat g^{-1} = \frac{1}{1+|z|^2} \left(
\begin{array}{cc}
 1-|z|^2 & 2 \bar z \\
 2z  & -1+|z|^2
\end{array}
\right),
\end{split}\end{align} 
when the north pole goes to the origin, and
\begin{align}\begin{split}
\hat g &=  \frac{1}{(1+|z|^2)^{1/2}}\left(
\begin{array}{cc}
\phantom{-}1 & z\\
-\bar z   & 1
\end{array}
\right),  \\
\hat q &= -\hat g \hat\sigma^z \hat g^{-1} = \frac{1}{1+|z|^2} \left(
\begin{array}{cc}
 -1+|z|^2 & 2 z \\
 2\bar z  & 1-|z|^2
\end{array}
\right)
\end{split}\end{align} 
for the opposite case. The $\hat q$-matrix here agrees with Eq.~(\ref{eq:q_sphere}) under the stereographic projection~(\ref{eq:n_via_z}). On the other hand, the functional form of $\hat g$ is different from the original definition~(\ref{eq:g_def}). The rule is that one considers any two matrices $\hat g$ and $\hat g'$ related by $\hat g' = \hat g e^{-i \psi \hat\sigma^z/2}$ to define the same element of the coset space. With this remark both definitions, (\ref{eq:g_def}) and (\ref{eq:g_def_z}), are equivalent since they correspond to the same density and $\hat q$-matrices.

The coherent state related to the above $\hat g$-matrix is defined in the same fashion as before,
\begin{align}\begin{split}
| g\rangle = \hat g |0\rangle \equiv \frac{\exp{(z \hat\sigma^-/2)} |0\rangle}{(1+|z|^2)^{1/2}} = 
\frac{|0\rangle  + z|1\rangle}{(1+|z|^2)^{1/2}}.
\end{split}\end{align} 
This agrees with the definition of the normalized coherent states in Ref.~\cite{Stone:2000}.
For completeness, we note that the resolution of identity takes the form
\begin{align}\begin{split}
\label{eq:I_res_z}
\int_{\mathds{C}} \mu(g) |g\rangle \langle g| = \sum_{s = 0,1}|s\rangle \langle s|, \quad \mu(g) = \frac{2}{\pi} \frac{dx dy}{(1+|z|^2)^2}
\end{split}\end{align}
in complex coordinates $ z = x + iy$.

\subsection{Berry Phase}\label{app:spin_path_berry}

To expand the Berry-phase term $S_I$ of Eq.~\eqref{eq:PathI_A} into coordinate representations, we use that $|0\rangle \hspace{-0.6mm}\langle 0| = (\mathds{1}  + \hat\sigma^z)/2$ together with the definition of spin coherent states to obtain
\begin{align}\begin{split}\label{eq:Berry_action}
 S_I &= \int_0^T dt\; \langle g| \partial_t  |g\rangle = \sum_i \int_0^T dt\; \langle 0| \hat g_i^{-1}  (\partial_t \hat g_i ) |0 \rangle  \\
 &= \frac 12 \sum_i \int_0^T dt\; \operatorname{Tr} \left[(\mathds{1} + \hat\sigma^z) \hat g_i^{-1} \partial_t \hat g_i\right] \\
 &= \frac 12 \sum_{i=1}^N \int_0^T dt\; \operatorname{Tr} \left[\hat\sigma^z \hat g_i^{-1} \partial_t \hat g_i\right].
\end{split}\end{align}
The first term of the expression in the second line is a boundary term evaluating to zero,
\begin{align}\begin{split}
\int_0^T {\rm tr} ( \hat g_i^{-1} \partial_t \hat g_i ) dt &= \int_0^T dt\, \frac{d}{dt} \ln {\rm det} \, \hat g_i = 0, 
\end{split}\end{align} 
since by the definition of unitary groups we have ${\rm det}\, g_i = 1$. The Berry phase written in the invariant form~(\ref{eq:Berry_action}) is a convenient starting point to derive specific coordinate representations. Using the explicit expressions for $g$ in either spherical or complex coordinates, one finds
\begin{align}\begin{split}\label{eq:S_Berry}
S_I = \frac 12 \sum_i \int_0^T dt\,\frac{\dot z_i \bar z_i - z_i {\dot{\bar{z_i}}}}{1 + |z_i|^2} 
=  - i \sum_i \int_0^T dt\, \dot\phi_i \cos\theta_i.
\end{split}\end{align}

\subsection{Hamiltonian}\label{app:spin_path_H}

To show how $S_H$ emerges, we start from the driving Hamiltonian.
We use the density matrix $\hat\rho = |g\rangle \langle g|$ defined in~(\ref{eq:rho_def}) to get
\begin{align*}
\langle g| \hat H_{D}(t) |g\rangle =- \frac{1}{2} \sum_i \Delta_i \operatorname{Tr} ( \hat\sigma^x  \hat q_{i}) = - \sum_i \Delta_i n_i^{x}.
 \end{align*}
The projection $n_i^{x}(t)$ can be now expressed either in terms of spherical angles (\ref{eq:q_sphere}), or complex coordinates $z_i$~(\ref{eq:n_via_z}), depending on the choice of parameterization of the Bloch sphere. Similarly, for the Ising Hamiltonian one finds
\begin{align}\begin{split}
\langle g| \hat H_{P}(t) |g\rangle &= - \sum_{i<j} J_{ij}  \operatorname{Tr} (\hat\sigma^z \hat \rho^{(i)})  \operatorname{Tr} (\hat\sigma^z \hat \rho^{(j)}) \\
&= -\frac 1 4 \sum_{i<j} J_{ij} \operatorname{Tr} ( \hat\sigma^z \hat q_i) \operatorname{Tr} ( \hat\sigma^z \hat q_j) \\
&= -\sum_{i<j} J_{ij} n_i^z n_j^z. 
\end{split}\end{align}
The linear combination of the two above pieces finally gives the action $S_H$.

\subsection{Mean-Field Equations as Saddle Point}

Here we use the least-action principle to derive the mean-field equations from the action $S = S_I + S_H$. One possible way is to accomplish this directly by using some coordinate system, say the complex coordinates from the stereographic projection. It is, however, instructive to also derive the equations of motion in a coordinate-free manner. 

To simplify the discussion, we consider a single spin rotating in the arbitrary magnetic field $\boldsymbol{B} = (B^x, B^y, B^z)^T$. This problem is described by an action $S$ with
\begin{align}\begin{split}
S_H &= - \frac  i 4 \int_0^T dt \operatorname{Tr} \left[B^\alpha \hat\sigma^\alpha \hat q\right] = -\frac i 2 \int_0^T dt\; \boldsymbol{B} \cdot \boldsymbol{n}.
\end{split}\end{align}
Again, $\hat q$ is the traceless part of the density matrix for each spin, see~(\ref{eq:q_sphere}).
Consider first the variation of the Berry phase, $S_I$, given by Eq.~(\ref{eq:Berry_action}). Let $\delta \hat g$ be a variation of $\hat g$. Since $\left(\delta \hat g\right) \hat g^{-1} + \hat g \delta \hat g^{-1} = 0$, hence
\begin{align}\begin{split}
\delta \hat g^{-1} = - \hat g^{-1} \delta \hat g \hat g^{-1}.
\end{split}\end{align}
A similar relation holds for $\partial_t g^{-1}$. Equipped with these relations we find for the variation of the Berry phase
\begin{align}
\delta S_I = -  \frac 12 \int dt \operatorname{Tr} (\hat g^{-1} \partial_t \hat q \, \delta \hat g ) .
\end{align}
Since $\delta g$ is an arbitrary unitary matrix, we find
\begin{align}\begin{split}
\hat g \frac{\delta S_I}{\delta \hat g} = - \frac 12 \partial_t \hat q.
\end{split}\end{align}
To get the variation of $S_H$ one has to proceed along the same lines. The result is
\begin{align}\begin{split}
\delta S_H = - \frac i 4 \int  dt\; B^\alpha \operatorname{Tr}\left[ \hat g^{-1} [\hat q, \hat\sigma^\alpha] \delta \hat g\right],
\end{split}\end{align} 
which yields
\begin{align}\begin{split}
\hat g \frac{\delta S_H}{\delta \hat g} = -\frac i 4  [\hat q, B^\alpha \hat\sigma^\alpha].
\end{split}\end{align}
Thus the saddle-point equations of motion are 
\begin{align}\begin{split}
i\partial_t \hat q = \frac 12 [\hat q, B^\alpha \hat\sigma^\alpha].
\end{split}\end{align}
Expanding $\hat q = n^\alpha \hat\sigma^\alpha$ and using the commutation relations $[\hat\sigma^\alpha, \hat\sigma^\beta] = 2i \epsilon_{\alpha\beta\gamma} \hat\sigma^\gamma$, the saddle-point equations can be rephrased as
\begin{align}\begin{split}
\partial_t \boldsymbol{n} = \boldsymbol{n} \times \boldsymbol{B},
\end{split}\end{align}
which is the Larmor precession of a spin in the magnetic field $B$.

The generalization to a multi-spin problem is now trivial.
Each spin is rotating in the effective magnetic field
\begin{align}\begin{split}
\label{eq:effective_field}
\boldsymbol{B}_i  &= 2 (1-s) \Delta_i \boldsymbol{\hat{e}}_x + 2 s m_i \boldsymbol{\hat{e}}_z,
\end{split}\end{align}
where  $m_i$ was defined in Eq.~\eqref{eq:magnetization}, and the equations of motion remain the same as above, $\partial_t \boldsymbol{n}_i = \boldsymbol{n}_i\times \boldsymbol{B}_i$.

For completeness, we now also give the derivation of the saddle-point equations complex-coordinate representation, Eqs.~\eqref{eq:eq_of_motion_z}. As shown in the main text, the starting point is the action in the form of Eq.~\eqref{eq:S_IH_z}. Neglecting boundary terms, the variation of $S_I$ then becomes
\begin{align}
    \begin{split}
        \delta S_I  = \sum_i \int dt \frac{\dot z_i \delta\bar z_i - \dot{\bar{z_i}} \delta z_i}{\left(1 + |z_i|^2\right)^{2}}.
    \end{split}
\end{align}
Together with the variation of the Hamiltonian part, one thus recovers Eqs.~\eqref{eq:eq_of_motion_z}.

\subsection{Derivation of the Action for Fluctuations}\label{subsec:A_flucs}

To obtain Eq.~(\ref{eq:S_fluctuations}) one should again start from Eq.~(\ref{eq:S_IH_z}) and substitute $z_i \to z_i + \delta z_i$ with
the variation $\delta z_i$ given by Eq.~(\ref{eq:delta_zi}).
On expanding in $\eta_i$, the linear terms will vanish, provided
the saddle-point equations~(\ref{eq:eq_of_motion_z}) are satisfied. Expanding to second order in the $\eta_i$ produces the Gaussian action of fluctuations~(\ref{eq:S_fluctuations}), where the diagonal elements of the matrices $ A$ and $ B$ are the same as found in Ref.~\cite{Stone:2000},
\begin{align}\begin{split}\label{eq:A_B_diag}
A_{ii} &= \frac 12 \frac{\partial}{\partial\bar z_i} (1 + |z_i|^2)^2\frac{\partial H}{\partial z_i} + ( z_i \leftrightarrow \bar z_i), \\
B_{ii} &=  \frac{\partial}{\partial\bar z_i} (1 + |z_i|^2)^2\frac{\partial H}{\partial \bar z_i}.
\end{split}\end{align}
while we find similar expressions for the off-diagonal elements,
\begin{align}\begin{split}\label{eq:A_B_off_diag}
A_{ij} &= (1 + |z_i|^2) (1 + |z_j|^2) \frac{\partial^2 H}{\partial \bar z_i \partial z_j}, \\
B_{ij} &=  (1 + |z_i|^2) (1 + |z_j|^2) \frac{\partial^2 H}{\partial \bar z_i \partial \bar z_j}.
\end{split}\end{align}
From these, one can see that $ A$ is Hermitian and $ B$ is symmetric. For the specific Hamiltonian~(\ref{eq:S_H_n_s}),
the above general expressions for the matrix elements then reduce to Eqs.~(\ref{eq:AB_diag}) and~(\ref{eq:AB_off-diag}).

Below we comment on some technical details used to derive the above expressions. We start from the Berry-phase contribution which produces the time-derivative term in the action~(\ref{eq:S_fluctuations}). To simplify the discussion, we assume that the final bit string has $\operatorname{sign}(n_i^z(T)) = 1$ for all $i$.
Substituting $\hat g_i' = \hat  g_i \tilde g_i e^{- i \psi \hat\sigma^z}$ into the Berry term~(\ref{eq:Berry_action}), one finds that the action is split into three terms,
\begin{align}\begin{split}
\label{eq:S_I_3}
S_I &= \frac 12  \sum_i \int_0^T\Bigl(\operatorname{Tr} \left(\hat\sigma^z e^{i \psi \hat\sigma^z} \partial_t e^{- i \psi \hat\sigma^z}\right) \\
&+\operatorname{Tr} \left(\hat\sigma^z \tilde g_i^{-1} \partial_t \tilde g_i\right) +\operatorname{Tr}\left(\tilde q_i g^{-1} \partial_t g \right)\Bigr),
\end{split}\end{align}
where we have defined $\tilde q_i = \tilde g_i \hat\sigma^z \tilde g_i^{-1}$. 
The first term evaluates to the boundary contribution 
\begin{equation}
 - i \sum_i \int_0^T  dt\, \partial_t \psi  = - i \sum_i [ \psi(T) - \psi(0)],
\end{equation}
which is zero since the fluctuations are absent at the boundaries, $\eta_i(0) = \eta_i(T)=0$. To obtain the contribution to the fluctuation action, one has to go to second order in the $\eta_i$. Then the second term in Eq.~(\ref{eq:S_I_3}) is again the Berry phase (\ref{eq:S_Berry}), yet now evaluated for $\tilde g_i$, which to this order becomes 
\begin{align}\begin{split}
\operatorname{Tr} (\hat\sigma^z \tilde g_i^{-1} \partial_t \tilde g_i) =  \bar \eta_i \partial_t \eta_i - \eta_i \partial_t \bar\eta_i .
\end{split}\end{align}
Written in matrix form, it reproduces the time-derivative term in the action~(\ref{eq:S_fluctuations}).  
To simplify the final term of the action (\ref{eq:S_I_3}), we note that $\tilde q_i$ is of the same form as the $\hat q$-matrix~(\ref{eq:g_def_q}), with $z$ replaced by $\eta_i$. When the latter are small, we find
\begin{equation}
\tilde q_i = \left(
\begin{array}{cc}
1 - 2|\eta_i|^2 & 2\bar\eta_i \\
2\eta_i  & - 1 + 2|\eta_i|^2
\end{array}
\right) + O(\eta_i^3), 
\end{equation}
which in turn generates the following second-order contribution to the action~(\ref{eq:S_I_3}):
\begin{align}
\label{eq:S_I_2}
S_I^{(2)} &\overset{\phantom{(\ref{eq:eq_of_motion_z})}}{=} \phantom{i}\sum_i \int_0^T dt\, |\eta_i|^2 \frac{(z_i {\dot{\bar{z_i}}} - \dot z_i \bar z_i)}{1+ |z_i|^2} \\
&\overset{(\ref{eq:eq_of_motion_z})}{=} 
i \sum_i \int_0^T dt\, |\eta_i|^2 (1+|z_i|^2) \left( \bar z_i \frac{\partial H}{\partial \bar z_i} +  z_i \frac{\partial H}{\partial z_i} \right),
\nonumber
\end{align}
where have also used the equations of motion~(\ref{eq:eq_of_motion_z}).

The variation of the Hamiltonian part of the action, $S_H$ in Eq.~(\ref{eq:S_IH_z}), is straightforward. On taking into account that to linear order, $\delta z_i = (1 + |z_i|^2) \eta_i$, one then arrives at the relations (\ref{eq:A_B_diag}) and (\ref{eq:A_B_off_diag}). The difference in the analytic expressions for the diagonal and off-diagonal elements stems from Eq.~(\ref{eq:S_I_2}), which contributes only to the
diagonal entries.

\vspace{0.5cm}

\section{Transfer matrix}
\label{sec:M_matrix}

In this Appendix, we summarize some basic facts on the transfer-matrix technique that was used in section~\ref{subsec:A_flucs}. At each time $t$, the transfer matrix $\boldsymbol{M}(t)$ can be written in its canonical form~\cite{Beenakker:1997},
\begin{equation}
\label{eq:M_decomposition}
M = \left(
\begin{array}{cc}
U' &  \\  & U^\dagger
\end{array}
\right)
\left(
\begin{array}{cc}
\cosh \boldsymbol{\lambda} & \sinh \boldsymbol{\lambda} \\  \sinh \boldsymbol{\lambda} & \cosh \boldsymbol{\lambda}
\end{array}
\right)
\left(
\begin{array}{cc}
V &  \\  & V'^\dagger
\end{array}
\right).
\end{equation} 
Here $\boldsymbol{\lambda} = {\rm diag}(\lambda_1, \lambda_2,...,\lambda_N)$ is the set of so called positive Lyapunov exponents, while $U,U',V,V'\in U(N)$ are unitary matrices. The time-dependence of these quantities is suppressed for brevity. Physically, the role of the unitaries is to rotate an initial basis of
in- and outgoing scattering states into 
a preferred basis, where the scattering occurs pairwise among right and left eigenmodes, which in turn are characterized by the corresponding Lyapunov exponents~$\lambda_l$. The block structure of the decomposition~(\ref{eq:M_decomposition}) matches the block form of the matrix~$\hat\tau_3$, see Eq.~(\ref{eq:H_t}), such that the law of `flux-conservation'~(\ref{eq:flux_conservation}) holds. 

At time $t=0$ scattering is absent, thus $\boldsymbol{M}(0)=0$. 
A special situation discussed in the main body of the paper is the so-called \textit{reflectionless} potential, when
all $\lambda_l(T)=0$. In this case, the transfer matrix $\boldsymbol{M}(T)$ is block-diagonal, such that $\left[\boldsymbol{M}_0(T), \hat\tau_3\right] = 0$. For the problem at hand, such reflectionless scattering potentials are realized by the effective Hamiltonian of paramagnons~(\ref{eq:H_t}) whenever the classical spin trajectories converge to the final bitstring $\boldsymbol{\sigma}_*$ exactly, i.e.~one has $n^{x,y}_i(T)=0$ for each spin (up to the numerical precision). We are not aware of a simple explanation of this remarkable fact.

The time dependence of the Green function under such conditions is simplified to $\boldsymbol{g}(t) = \boldsymbol{M}(t) \hat\tau_3 \boldsymbol{M}^{-1}(t)$.
Indeed, since in this case the matrices $\boldsymbol{M}(T)$ and $\hat\tau_3$ commute at $t=T$, the evolution brings $\boldsymbol{g}(T)$ back to $\hat\tau_3$, and both the initial and final values of $\boldsymbol{g}(t)$ are in accord with the boundary condition~(\ref{eq:bc_Q_matrix}) for the spin path integral.

\bibliography{refs}

\end{document}